\documentclass{jfm}
\usepackage{float}
\usepackage{graphicx}
\usepackage{array,multirow}
\usepackage{gensymb}

\begin{document}

\newtheorem{lemma}{Lemma}
\newtheorem{corollary}{Corollary}

\shorttitle{Lump Collision} 
\shortauthor{Masnadi and Duncan} 

\title{Observations of gravity-capillary lump interactions}

\author
 {
 Naeem Masnadi\aff{1}
 \corresp{\email{nmasnadi@gmail.com}},
  \and 
James H. Duncan\aff{1}
  }

\affiliation
{
\aff{1}
Department of Mechanical Engineering, University of Maryland, College Park, MD 20742, USA
}

\maketitle

\begin{abstract}

In this experimental study, we investigate the interaction of gravity-capillary solitary waves generated by two surface pressure sources moving side by side at constant speed. The nonlinear response of a water surface to a single source moving at a speed just below the minimum phase speed of linear gravity-capillary waves in deep water ($c_{min}\approx23$ cm s$^{-1}$) consists of periodic generation of pairs of three-dimensional solitary waves (or lumps) in a V-shaped pattern downstream of the source. In the reference frame of the laboratory, these unsteady lumps propagate in a direction oblique to the motion of the source. In the present experiments, the strengths of the two sources are adjusted to produce nearly identical responses and the free surface deformations are visualized using photography-based techniques. The first lumps generated by the two sources move in intersecting directions that make a half angle of approximately $15\degree$ and collide in the center-plane between the sources. A steep depression is formed during the collision, but this depression quickly decreases in amplitude while radiating  small-amplitude radial waves. After the collision, a quasi-stable pattern is formed with several rows of localized depressions that are qualitatively similar to lumps but exhibit periodic amplitude oscillations, similar to a breather. The shape of the wave pattern and the period of oscillations depend strongly on the distance between the sources.

\end{abstract}

\section{Introduction}

A solitary wave is a localized disturbance of permanent form resulting from a perfect balance between opposing effects of dispersion and nonlinearity. A defining feature of soliton solitary waves (as are found in completely integrable model equation systems) is that they  interact elastically with other solitary waves, i.e. they remain unchanged after the interaction, except for a phase shift. 
Real solitary water waves exhibit both elastic and non elastic features during their interactions.  
In shallow water, two-dimensional gravity solitary waves bifurcate from the linear dispersion curve at zero wavenumber. The collision of real gravity solitary waves has been studied extensively \cite[for example, see][]{Craig2006,Yeh2014}.
In deep water, the minimum phase speed of linear gravity-capillary waves is known to be the bifurcation point of two-dimensional and three-dimensional solitary waves (the latter are usually referred to as `lumps' or `wavepacket solitary waves') \cite[]{LH1989,Kim2005,Milewski2005}. 
Some aspects of interactions of gravity-capillary solitary waves in deep water have been explored numerically.
\cite{Milewski2010} performed simulations of the Euler equations for head-on and overtaking collisions of two-dimensional gravity-capillary solitary waves. They found that head-on collisions were almost elastic with little radiation but for overtaking collisions the smaller wave can break up and transfer some energy to the larger wave. 
The dynamics of lumps during collision was studied by \cite{Akers2009,Akers2010} using model equations based on truncations of the Euler equations. Similary to the two-dimensional case, they found no visible interaction for head-on collisions but found that overtaking collisions typically result in the formation of one larger lump and a radiated wave field. Employing a different model, \cite{Milewski2012} only observed quasi-elastic interactions where both lumps survived, although they did not rule out the possibility of cases where only one lump remains  after the collision. To the best of our knowledge, there have been no experimental observations of collision of gravity-capillary lumps in the literature.

In this paper, we investigate the dynamics of the wave pattern produced by two pressure distributions moving side-by-side at a speed close to $c_{min}$ and report the first experimental observations of the collision of gravity-capillary lumps.
In clean water, the minimum phase speed of linear gravity-capillary waves is $c_{min}=23.13$ cm s$^{-1}$ which occurs at a wavelength of $\lambda_{min}=1.71$ cm. 
The nonlinear response of a water surface to a single surface pressure distribution moving at a speed very close to $c_{min}$ consists of periodic generation of depression lumps downstream of the disturbance in a V-shaped pattern. This unsteady response was first discovered by \cite{DiorioPRL,DiorioJFM} and named `state III'. These lumps are shed in pairs from the tips of the V-shaped pattern and exhibit characteristics similar to those of freely-propagating lumps of potential flow theory, first calculated by \cite{Parau2005}. As the lumps move away from the source, they decay due to viscous damping effects and closely follow the speed-amplitude relationship of steady lumps of potential flow theory \cite[]{MasnadiJFM}. 
In the reference frame of the laboratory, the travel direction of the lumps makes an angle of approximately $15\degree$ with the direction of motion of the source. This provides a scenario for using two jets moving side by side to generate two nearly identical lumps simultaneously and observing their collision at an oblique angle. 
This work might also be applicable to the case of two trucks or hovercraft moving side by side on a floating ice sheet, since the floating ice system also supports solitary waves bifurcating from linear dispersion relationship at a minimum phase speed found at finite wave number \cite[for example, see][]{IceSheet}.

\section{Experimental details}\label{sec:ExpDetails}

The experiments reported herein were performed in the same towing tank and with the same measurement techniques as those described in \cite{MasnadiThesis} and \cite{MasnadiJFM}.  A brief overview of these facilities and techniques are given below; the interested reader is referred to the original references for further details.

\begin{figure}
\begin{center}
  \includegraphics[width=4.0in]{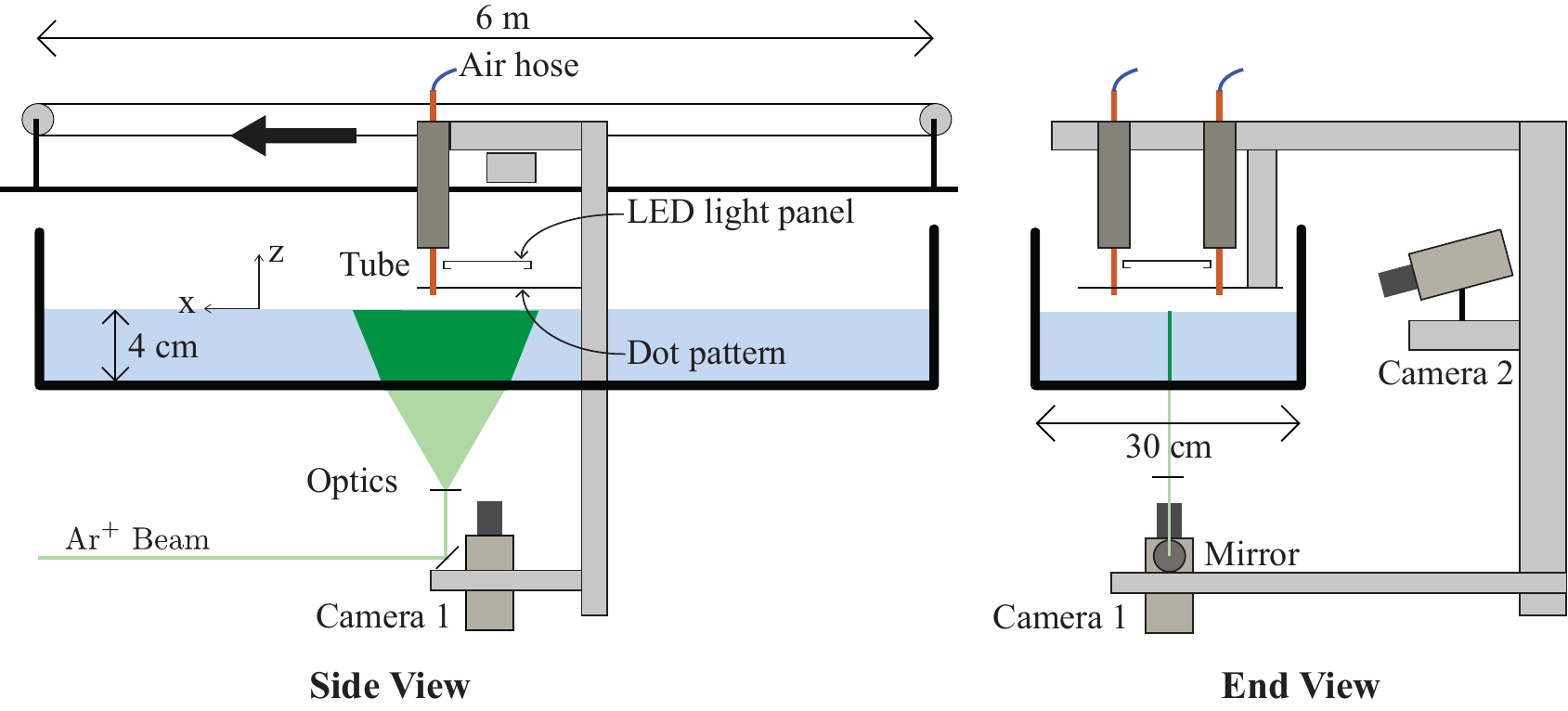}
\end{center}
\caption{Schematic drawing of the experimental set-up showing two views of the towing tank, tubes, and measurement systems.}
\label{fig:setup}
\end{figure}

The experiments are performed in a towing tank that is 6~m long, 30~cm wide and 7~cm deep (see figure \ref{fig:setup}). 
The set-up includes an instrument carriage that travels along the length of the tank. Two pressure disturbances are made on the water surface by blowing air through two vertically oriented tubes of inner diameter 2.5 mm. 
These tubes are attached to the instrument carriage and held in a plane whose normal is in the direction of the carriage motion. (The experiments in \cite{MasnadiJFM} were performed with only one of these tubes.) The bottom ends of the tubes are positioned at approximately 7~mm above the water surface.  The air flow in each tube is supplied by a completely independent system.   Since surface tension effects are important in this experiment, great care was taken to make sure that the water remained relatively surfactant-free during all experimental runs.

Qualitative observations of the main features of the surface response are visualized using a refraction-based technique. 
In this technique, a translucent surface with a pattern of randomly placed dots is attached to the carriage via a vertically oriented traverser and held parallel to and above the calm water surface. 
The distance between the dot pattern and the water surface can be adjusted using the traverser. 
A vertically oriented high-speed movie camera (camera 1 in figure \ref{fig:setup}) images the dot pattern through the transparent bottom of the tank and the water free surface. 
Due to refraction at the air-water interface, the dot pattern images become distorted when the surface is deformed, and it is possible to find the location of surface depressions by interpreting the images. 
When the dot pattern is far enough from the free surface, the dots appear as lines at places with high surface slope (e.g. between a crest and a trough), and the dot density increases at the trough of a depression. 
The lensing effect of the surface curvature also changes the local light intensity in the images.

A cinematic laser-induced fluorescence (LIF) method is used to measure the wave height along the center-line of the two tubes, see figure \ref{fig:setup}. 
In order to perform these measurements, Fluorescein dye is added to the water, and a thin sheet of light from an argon-ion laser is projected vertically onto the water surface from below. The light sheet is approximately 20~cm wide and 1~mm thick, and the plane of the light sheet is oriented in the direction of the carriage motion.   A second camera is mounted on the carriage (camera 2 in figure \ref{fig:setup}) and views the intersection of the light sheet and the free surface through the transparent side wall and from above the surface, looking down with an angle of $15\degree$ from the horizontal. The free surface shape along the laser sheet is found using a gradient-based edge detection method. It is estimated that the free surface can be located to within $\pm 1$~pixel ($\pm 0.02$~mm) with this LIF method.

The strength of free surface forcing is controlled by the airflow in the air-jet tubes and is denoted by the non-dimensional parameter $\epsilon=h_0/d$, where $h_0$ is the depth of the circular depression under the tube when the carriage is stationary and $d$ is the internal diameter of the tube.  The towing speed of the carriage is indicated by $\alpha=U/c_{min}$, where $U$ is the carriage speed. In all experiments reported herein, $\epsilon=0.3$ and $\alpha = 0.994$ (towing speed of 23 cm/s).  For these values of $\epsilon$ and $\alpha$, when only one pressure source is active, the response is in state III as defined by \cite{DiorioPRL}. State III includes periodic shedding of lumps from the tips of a V-shaped pattern behind the air jet. The distance between the air-jet tubes is scaled by $\lambda_{min}=1.71$~cm, and denoted by the non-dimensional parameter $\delta_{T}$. In the experiments presented, $\delta_{T}$ varies between 2.3 and 7.0.

\section{Results and discussion}\label{sec:Results}

\begin{figure}
\begin{center}
\begin{tabular}{ccc}
(\textit{a}) & (\textit{b}) & (\textit{c})\\
  \includegraphics[width=1.7in]{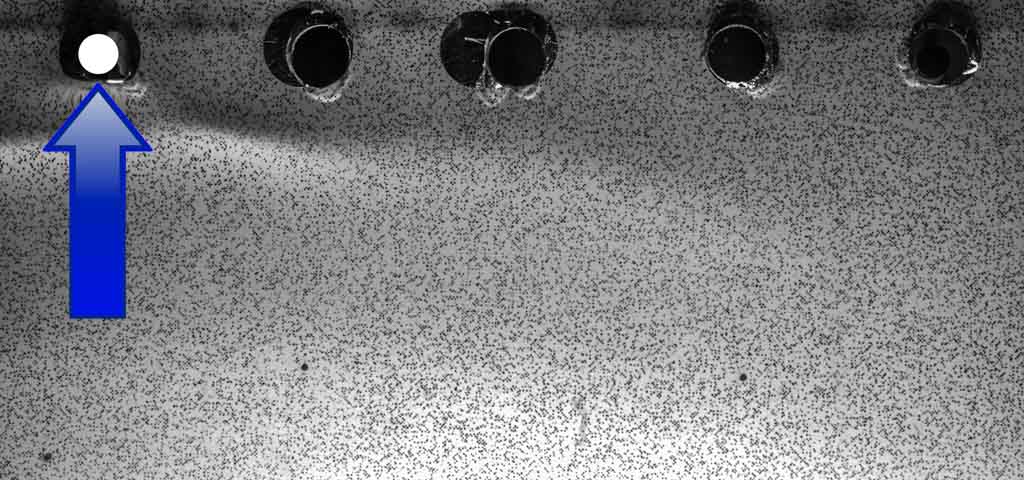}&
  \includegraphics[width=1.7in]{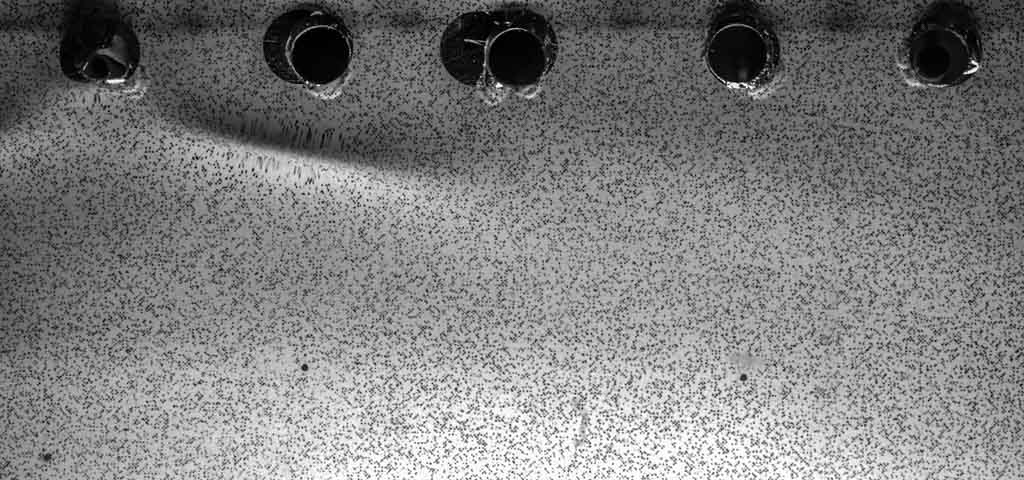}&
  \includegraphics[width=1.7in]{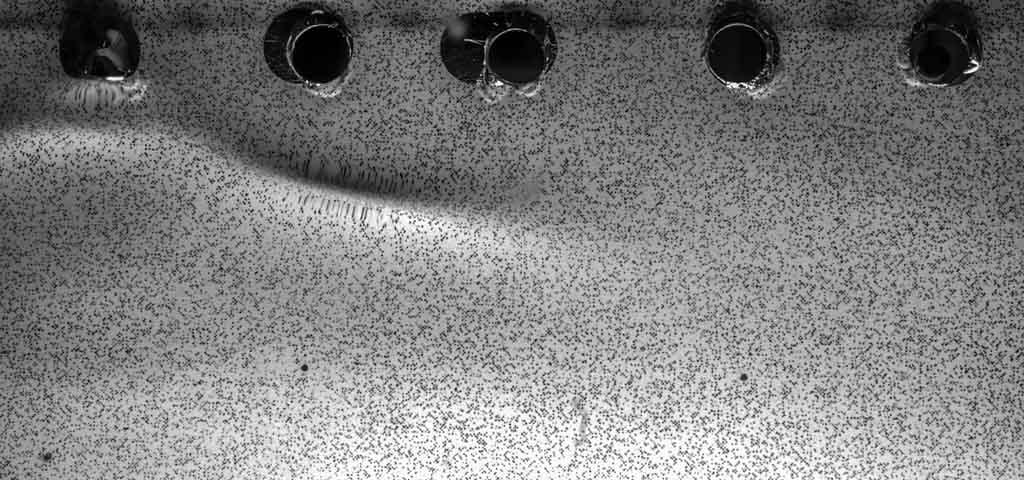}\\
(\textit{d}) & (\textit{e}) & (\textit{f})\\
  \includegraphics[width=1.7in]{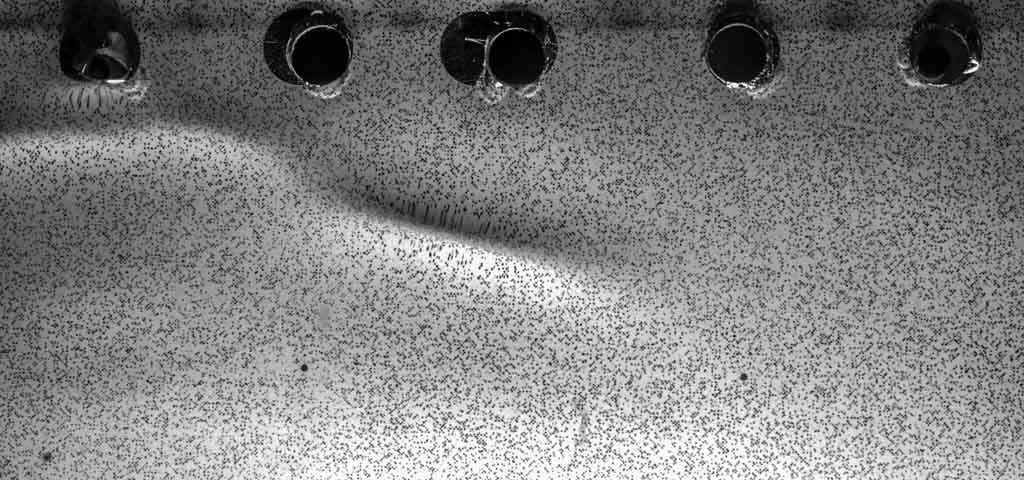}&
  \includegraphics[width=1.7in]{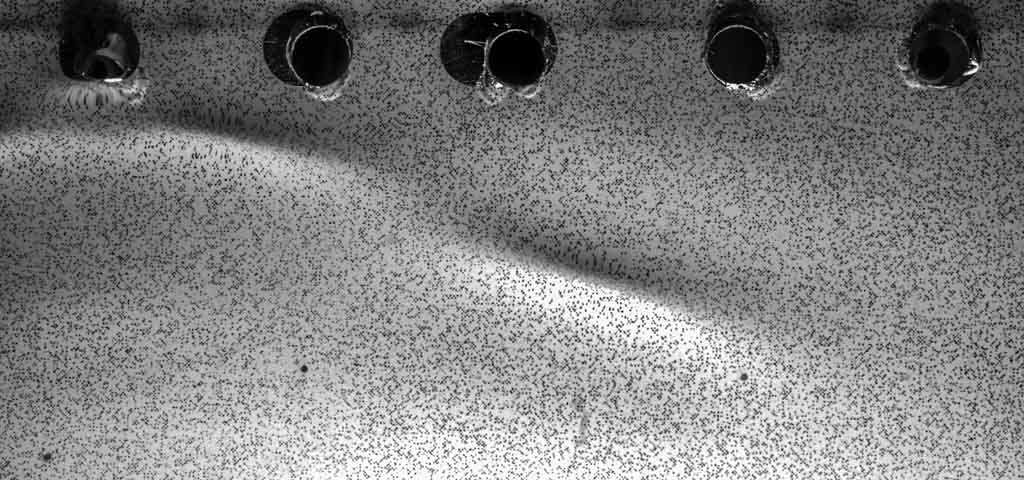}&
  \includegraphics[width=1.7in]{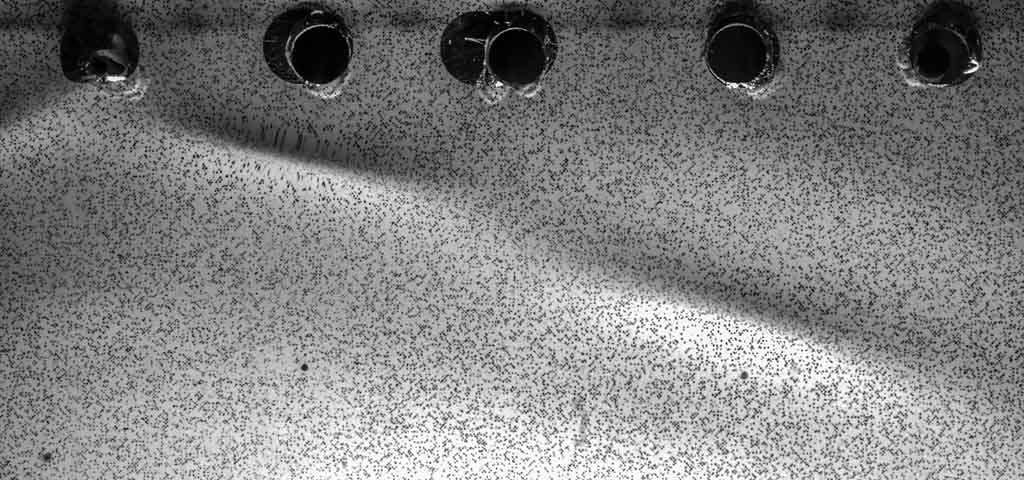}\\
\end{tabular}
\end{center}
\caption{A sequence of refraction images with one active air jet. The five dark circular regions along the top boundary of the images are holes in the mounting plate for the dot pattern; the air jet is located in the top-left hole,  which is marked by a white dot.  The arrow in (\textit{a}) indicates the towing direction of the source. Each image is approximately 10 cm wide in the physical plane. The time separation between images is 0.25 s.}
\label{fig:SingleSource}
\end{figure}

\begin{figure}
\begin{center}
\begin{tabular}{ccc}
(\textit{a}) & (\textit{b}) & (\textit{c})\\
  \includegraphics[width=1.7in]{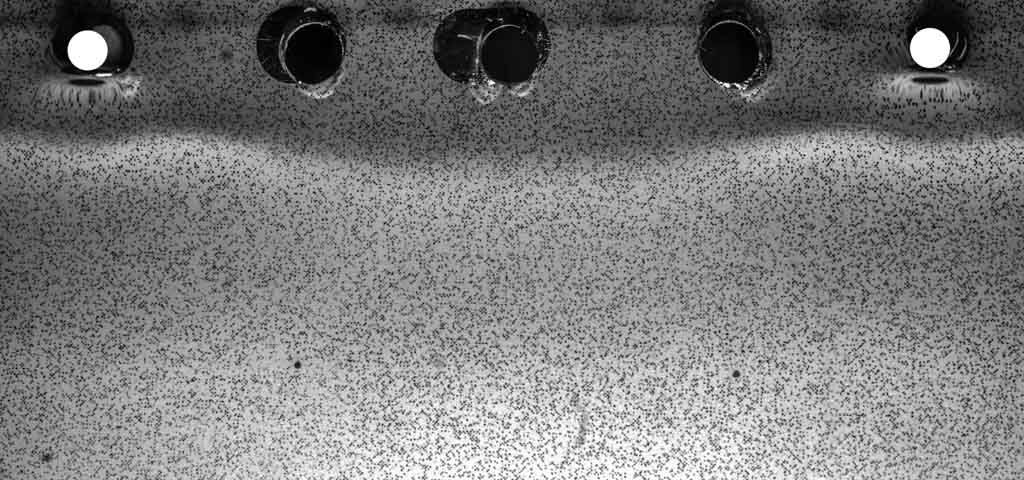}&
  \includegraphics[width=1.7in]{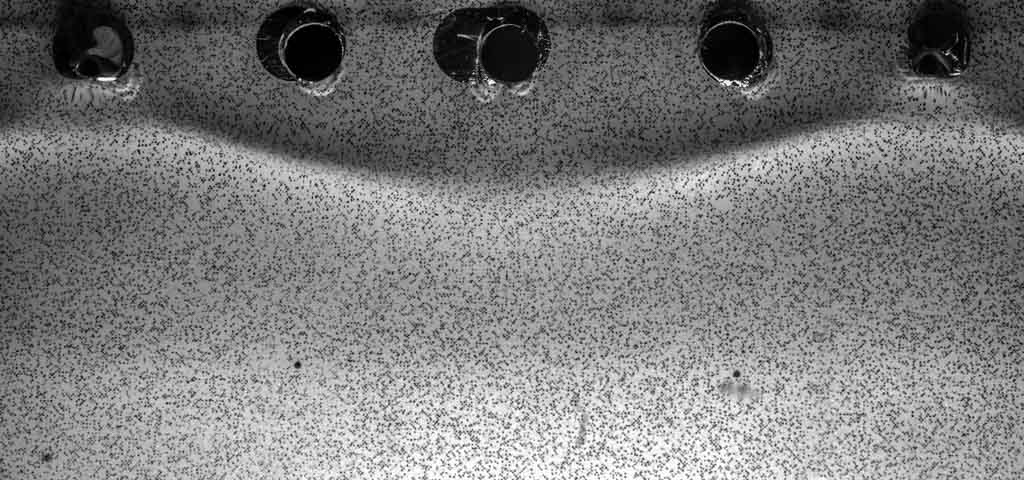}&
  \includegraphics[width=1.7in]{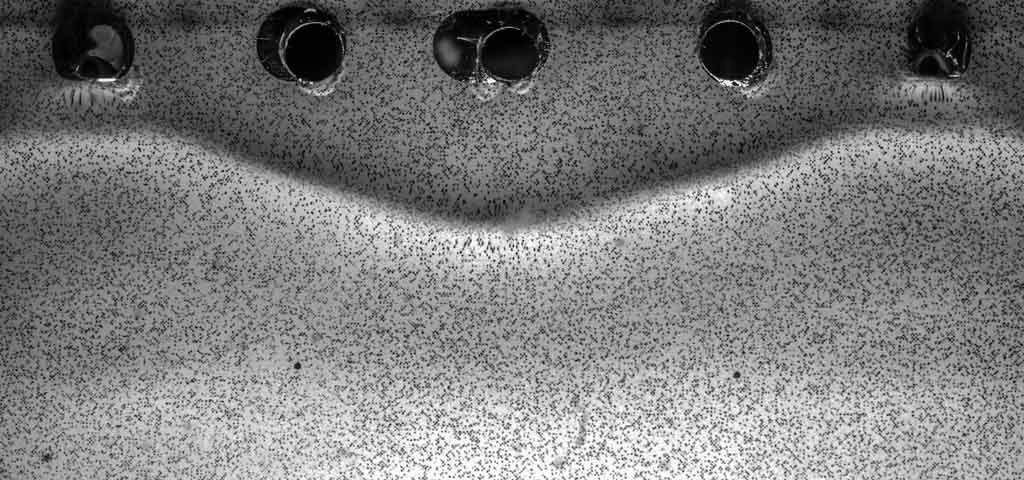}\\
(\textit{d}) & (\textit{e}) & (\textit{f})\\
  \includegraphics[width=1.7in]{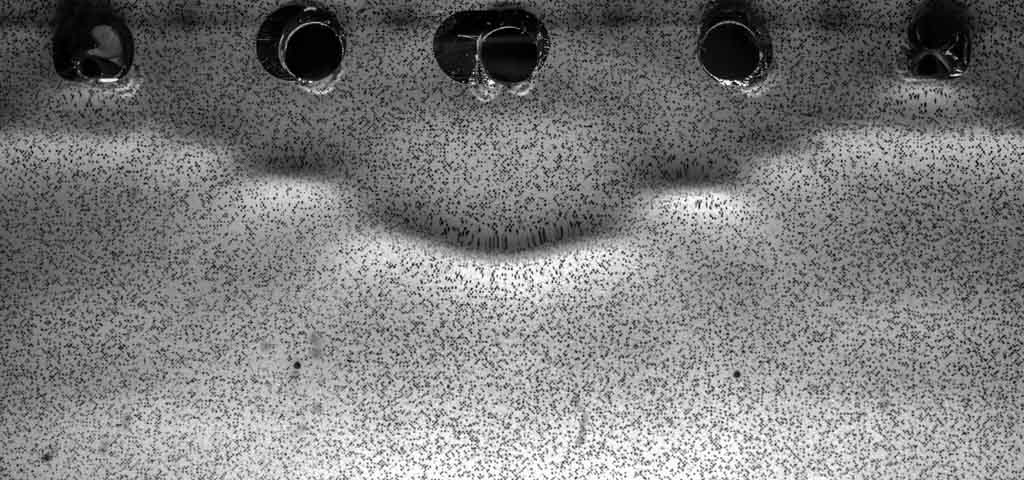}&
  \includegraphics[width=1.7in]{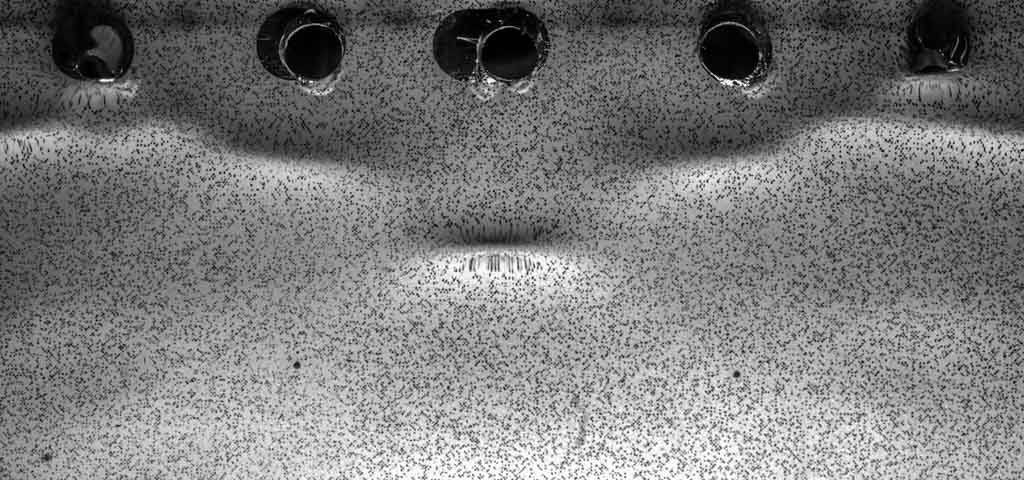}&
  \includegraphics[width=1.7in]{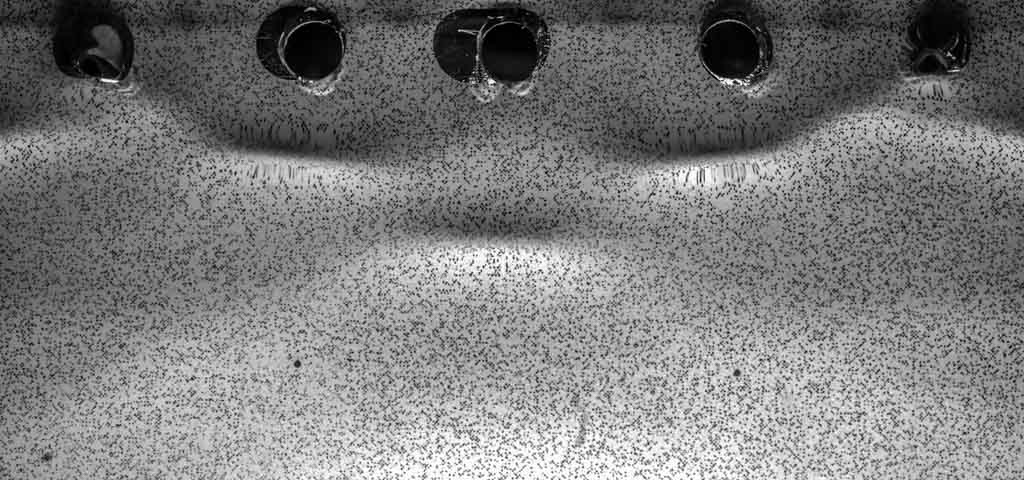}\\
\end{tabular}
\end{center}
\caption{A sequence of refraction images with two active air jets, $\delta_{T}=4.7$. The tubes are located in the leftmost and rightmost mounting holes at the top of the images. The time separation between images is 0.25 s.}
\label{fig:D8}
\end{figure}

In this section, the water surface deformation patterns produced by the two air jets are explored.  A qualitative description of the free surface response is first discussed using the raw refraction images. These images provide a full-field visualization of the free surface shape. This is followed by the quantitative LIF measurements of the surface height along the streamwise midline between the two air jets.

The behavior of the two-jet surface pattern is better understood by first comparing it during its formation with the pattern development for a single air jet.  
To this end, two sequences of six refraction images are shown in figures \ref{fig:SingleSource} and \ref{fig:D8} (also see movie 1 and movie 2 in supplementary files). 
In figure \ref{fig:SingleSource}, the surface deformation pattern is generated by a single air jet located near the top-left corner of each image, while in figure \ref{fig:D8}, there is an air jet in both the left and right top corners, separated by 8~cm in the physical plane ($\delta_{T}=4.7$). 
In both figures, the first images were obtained just after the air-jet tube reached constant speed. 
The images in each sequence are separated in time by 0.25~s. 
For the single air jet, the pattern develops as described qualitatively by \cite{DiorioPRL} and measured in detail by \cite{MasnadiJFM}. 
At this condition, a sequence of pairs of depression lumps is generated periodically at the ends of a small V-shaped pattern, and once formed, each lump propagates along an approximately straight line ray oriented at an angle of approximately $15\degree$ to the direction normal to the carriage motion. 
In the images in figure~\ref{fig:SingleSource}, only the right side of the symmetric pattern is shown. 
The long axis of the lump is nearly aligned with the rays.
The propagation speed of the lump along the ray is relatively slow in the images, but in the reference frame of the laboratory, the lump moves along the tank at a speed a little less than $c_{min}$  in the direction perpendicular to the long axis of the lump. 
The shedding period of these pairs of lumps was reported in \cite{MasnadiJFM} and found to decrease with increasing towing speed, reaching around 1 s as the towing speed approaches $c_{min}$. 
The shedding period is approximately 1 s for the parameters used in the current experiments.
The timing of the shedding of the first lump is fairly repeatable from run to run.

When both air jets are turned on (movie 2), the shedding of the first lumps from the two air jets is nearly synchronized, see the first two images in figure~\ref{fig:D8}. 
In the third image, the left edge of the lump from the right air-jet tube and the right edge of the lump from the left air-jet tube meet in the center-plane. As the lumps collide, a pattern of small-amplitude waves radiates radially away from the location of the collision; one of these radiated waves is barely visible in the top center of the fourth image. The radiated waves have small amplitude, but can be seen clearly in the movie image sequence corresponding to this figure (see supplementary files online). By the time of the fifth image, the radiation has stopped and a pattern of three rows of isolated depressions has formed with the first row of two depressions directly behind the air jets, the second row of two depressions 6~mm downstream and the third row, consisting of a single depression, 7~mm farther downstream and centered on the midline between the two air jets. The depressions in the second and third rows have become oriented with their long axes nearly perpendicular to the flow direction. No evidence of transmission or reflection of the original two lumps is found. While this basic pattern seems to be stable, the streamwise position of the depressions varies by a small amount in time and small-amplitude waves radiate from the middle depression periodically. Radiation events at the center-plane depression occur with a seemingly random periodicity.

\begin{figure}
\begin{center}
\begin{tabular}{ccc}
(\textit{a}) & (\textit{b}) & (\textit{c})\\
  \includegraphics[width=1.7in]{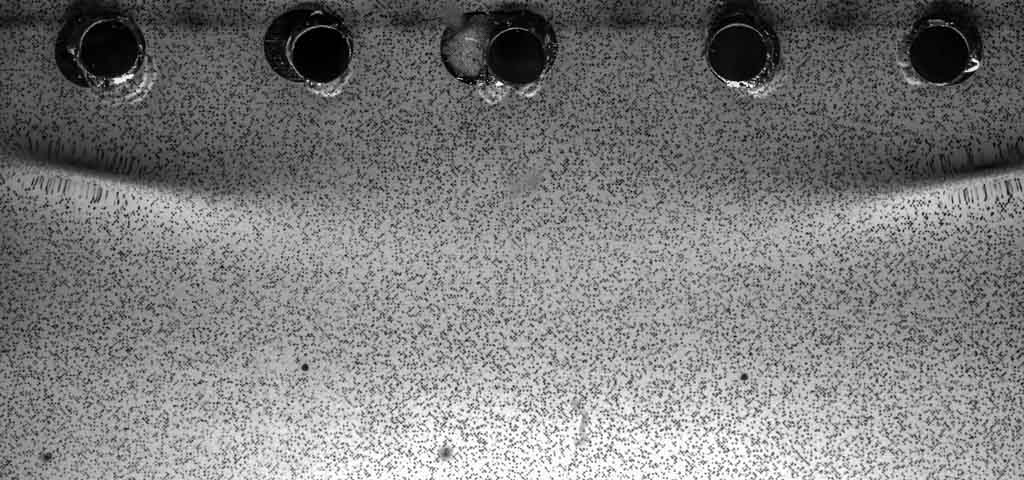}&
  \includegraphics[width=1.7in]{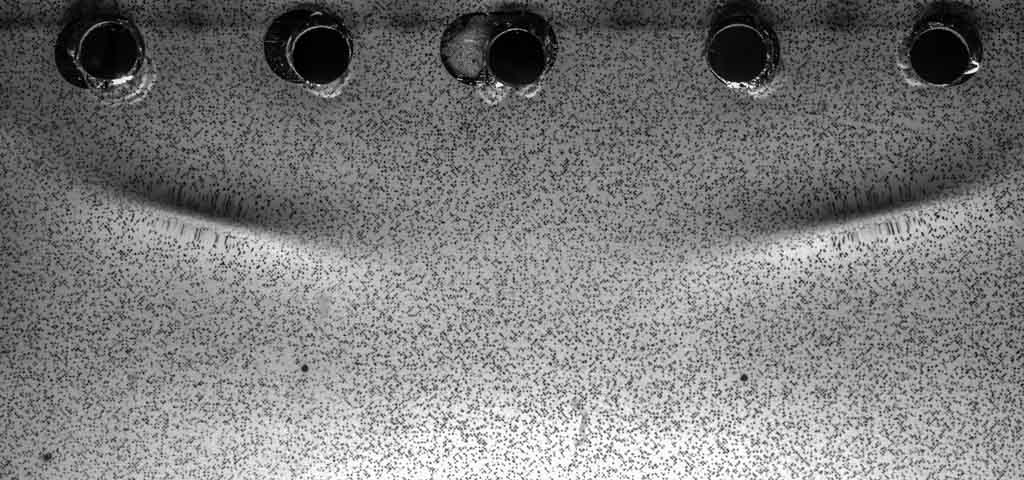}&
  \includegraphics[width=1.7in]{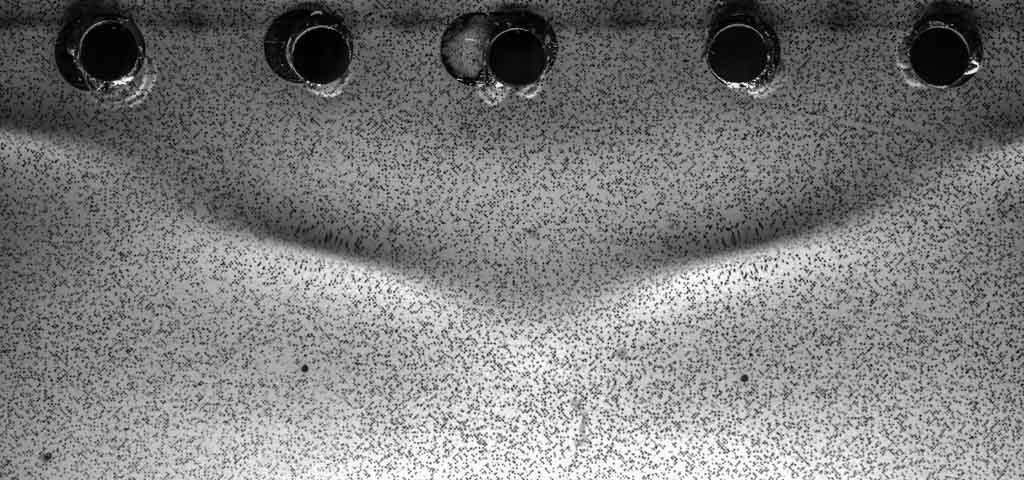}\\
(\textit{d}) & (\textit{e}) & (\textit{f})\\
  \includegraphics[width=1.7in]{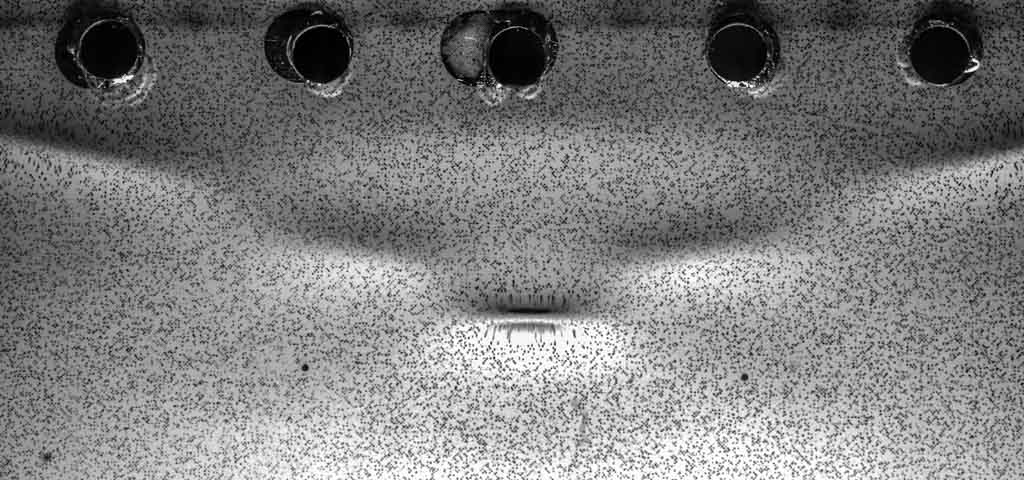}&
  \includegraphics[width=1.7in]{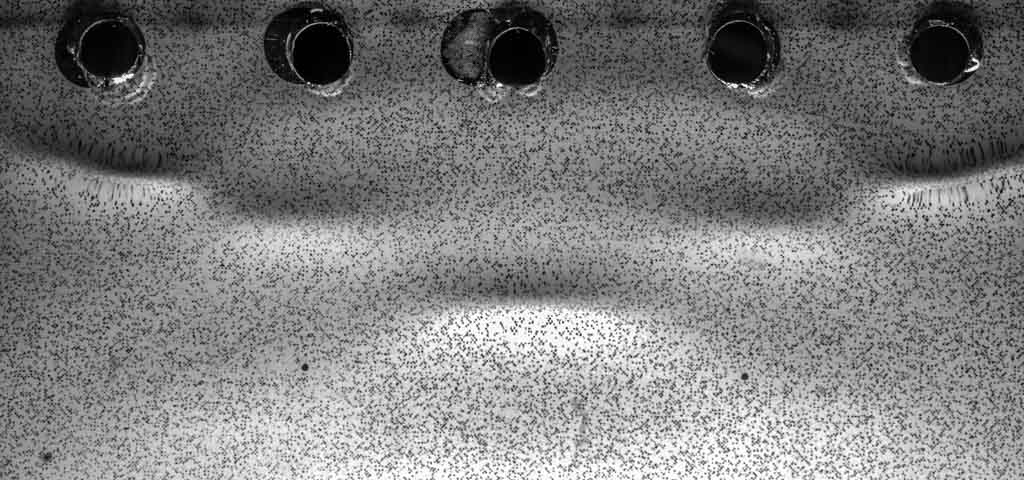}&
  \includegraphics[width=1.7in]{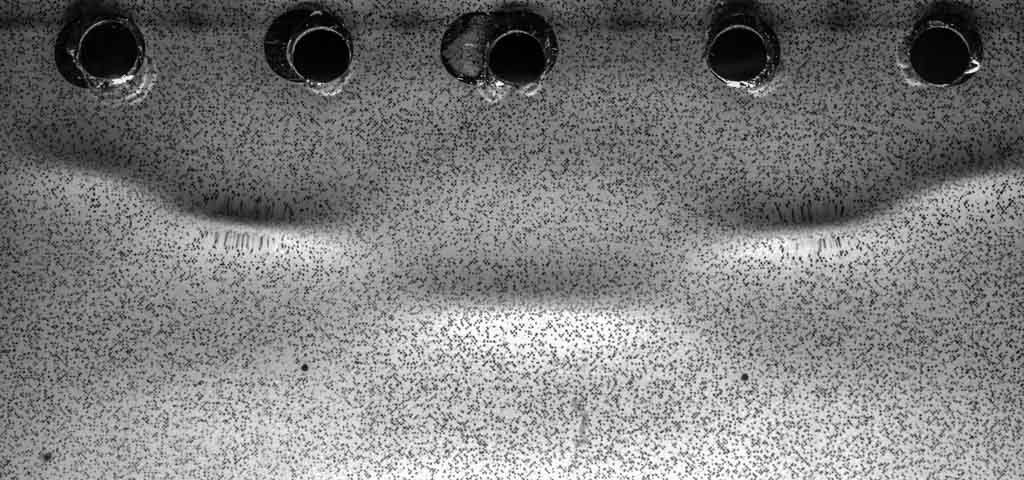}\\
\end{tabular}
\end{center}
\caption{A sequence of refraction images with two active air jets, $\delta_{T}=7.0$. The tubes are located outside the field of view, in mounting holes on the right and left sides of the images.  The time separation between images is 0.25 s.}
\label{fig:D12}
\end{figure}

\begin{figure}
\begin{center}
\begin{tabular}{cccc}
(\textit{a}) & (\textit{b}) & (\textit{c}) & (\textit{d})\\
  \includegraphics[width=1.25in]{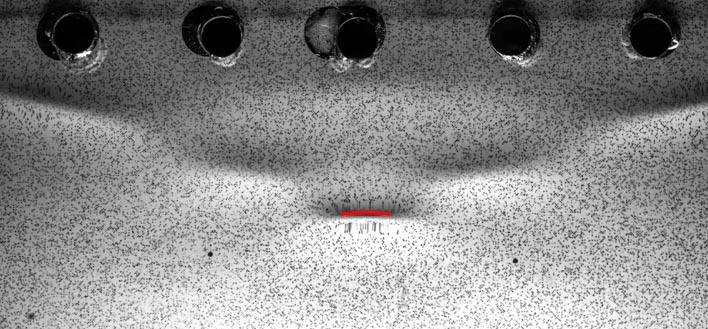}&
  \includegraphics[width=1.25in]{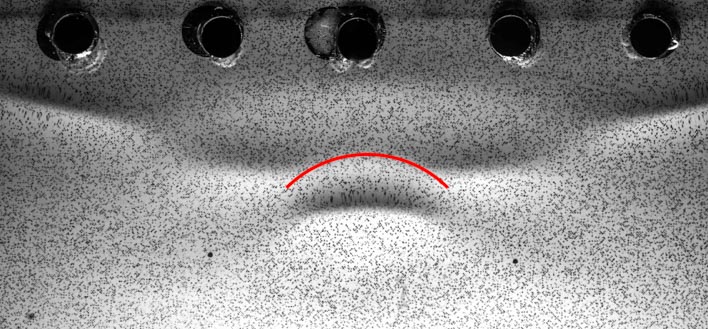}&
  \includegraphics[width=1.25in]{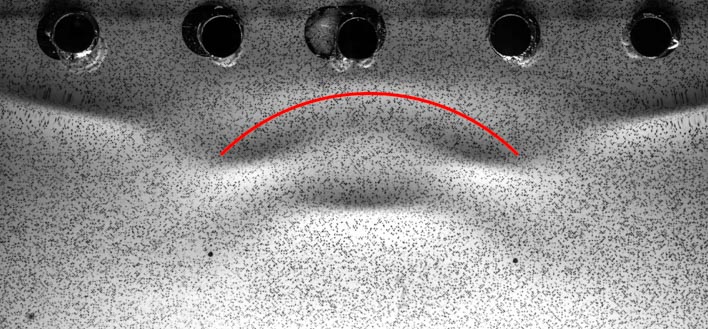}&
  \includegraphics[width=1.25in]{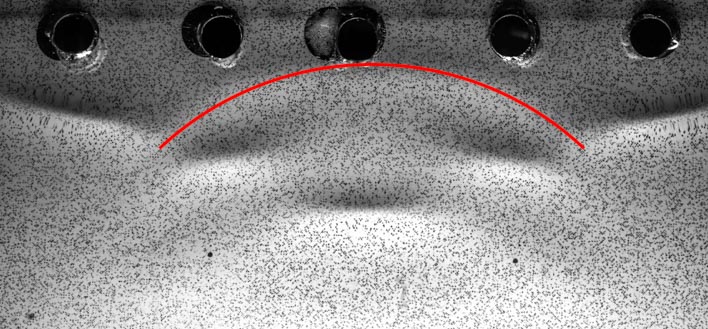}\\
\end{tabular}
\end{center}
\caption{Propagation of small-amplitude radial waves after the first collision for $\delta_{T}=7.0$.  The images are separated by 0.15 s.}
\label{fig:radiation}
\end{figure}

The collision event and the radiation of small-amplitude radial waves can be seen more clearly in figures \ref{fig:D12} and \ref{fig:radiation} where the tube separation is fixed at $\delta_{T}=7.0$ cm (also see movie 3, movie 4a, and movie  4b). In figures \ref{fig:D12} (\textit{a})-(\textit{c}), two isolated lumps approach the center-plane and collide in figure \ref{fig:D12}-(\textit{d}) and form a pattern with several rows of localized depressions oriented normal to the carriage motion. This pattern is relatively stable with the depth of the middle depression oscillating in a manner similar to the breathers described in \cite{Milewski2012}. A breather is a time-periodic localized structure, in this case a depression lump, with periodic amplitude modulations. A sequence of images just after the collision is shown in figure \ref{fig:radiation} with a time separation of 0.15 s between images. The propagation of a small-amplitude radial wave from the collision location is visible in these images. (The wavefront is denoted by red arcs.) From these images, the propagation speed of the wavefront is measured to be approximately 27~cm s$^{-1}$ in the reference frame of the laboratory.  This  speed corresponds to a wavelength of 0.75~cm on the capillary branch of the linear theory dispersion curve and 3.92~cm on the gravity branch. Of course, the burst of small-amplitude wave generation is likely to be composed of a spectrum of wavelengths, but the wavelength observed in the refraction images is about 1.5~cm. The radiation of linear waves during head-on and overtaking lump collisions was also observed in the calculations of \cite{Akers2009}.

\begin{figure}
\begin{center}
\begin{tabular}{ccc}
(\textit{a}) $\delta_{T}=2.3$ & (\textit{b}) $\delta_{T}=3.5$ & (\textit{c}) $\delta_{T}=4.7$\\
  \includegraphics[width=1.7in]{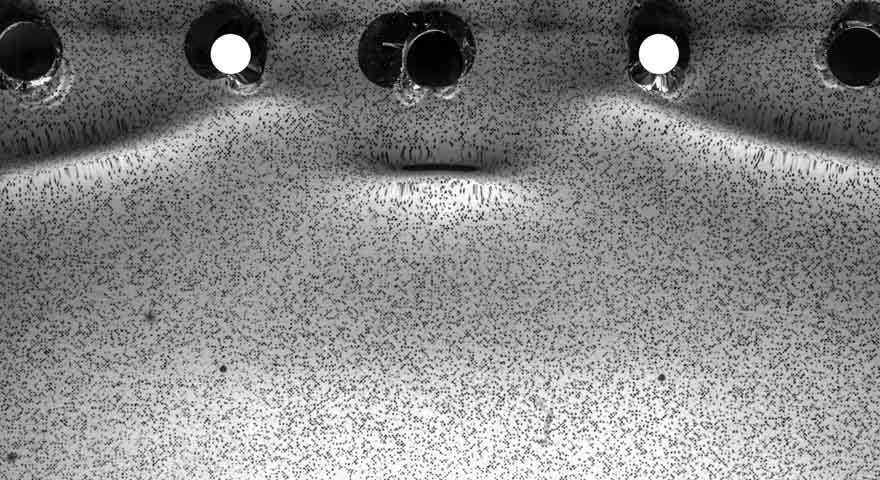}&
  \includegraphics[width=1.7in]{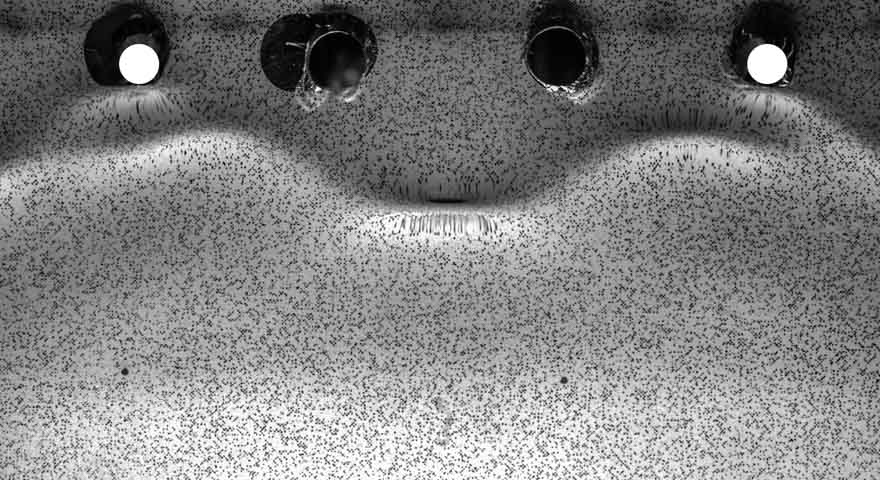}&
  \includegraphics[width=1.7in]{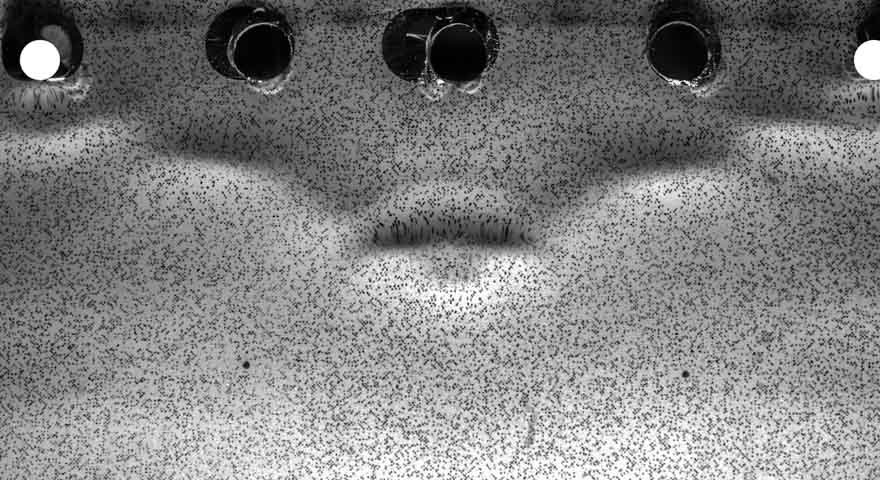}\\
\end{tabular}
\begin{tabular}{cc}
(\textit{d}) $\delta_{T}=5.8$ & (\textit{e}) $\delta_{T}=7.0$\\
  \includegraphics[width=1.7in]{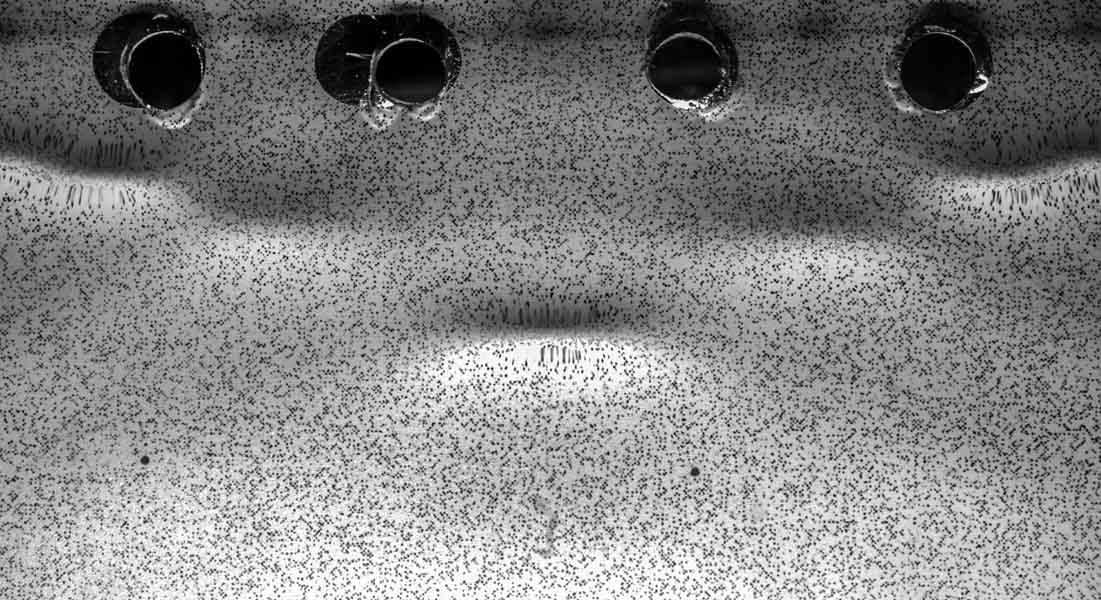}&
  \includegraphics[width=1.7in]{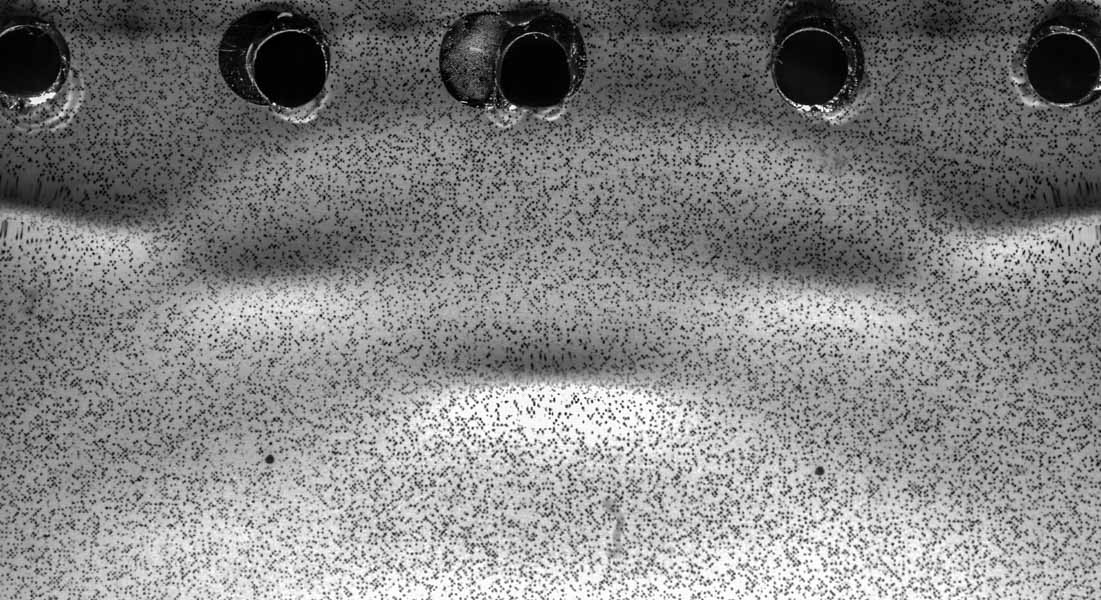}\\
\end{tabular}
\end{center}
\caption{Refraction images 0.2~s after the first burst for different tube separations.} 
\label{fig:SeveralSeparations}
\end{figure}

The distance between the two air-jet tubes has a strong influence on the surface deformation pattern. This pattern is shown in figure~\ref{fig:SeveralSeparations} for $\delta_{T}=$ 2.3, 3.5, 4.7, 5.8 and 7.0. The images are cropped so that the middle depression is in the middle of each image and the active air jets for the $\delta_{T}=$~5.8 and 7.0 cases are outside the field of view. For each condition, the deformation pattern at the instant 0.2 s after the first burst event  is shown. If the disturbance under and just behind the air-jet tubes is called the first row, then there are two rows of depressions for the $\delta_{T}=$~2.3 case and four rows of depressions for the $\delta_{T}=$ 7.0 case.  The oscillations of the breather mode of the pattern and the lump at the collision site are discussed along with the LIF surface profile measurements later in this section.

\begin{figure}
\begin{center}
\begin{tabular}{cc}
  (\textit{a}) & (\textit{b})\\
  \includegraphics[width=2.2in]{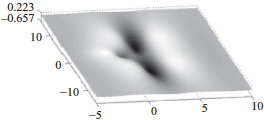}&
  \includegraphics[width=2.2in]{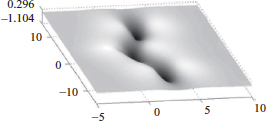}\\
\end{tabular}
\end{center}
\caption{Computed steady patterns with multiple lumps generated by a single source. From figure~7 in \cite{ChoJFM}. (\textit{a}) $\alpha=0.970$,  (\textit{b}) $\alpha=0.930$.}
\label{fig:ChoMultipleLumps}
\end{figure}

The stable pattern with several rows of depressions is qualitatively similar to the steady surface deformation patterns found by \cite{ChoJFM} for a single pressure source moving at a constant speed below $c_{min}$, see Figure~\ref{fig:ChoMultipleLumps}(\textit{a}) and (\textit{b}), which were taken from their paper. In these calculations, \cite{ChoJFM} looked for steady solutions to their model equation by using a numerical continuation method and found various response patterns according to the position of the solution along the bifurcation curve of maximum depression depth versus dimensionless source speed, $\alpha = U/c_{min}$.  In the presence of damping, the curve exhibits multiple turning points and the patterns shown in figure \ref{fig:ChoMultipleLumps} are obtained just  after the third (\textit{a}) and fourth (\textit{b}) turning points.  In these patterns, as in the present experiment there are rows of one and two depressions, which are oriented in the direction of the source motion.  These steady patterns were not found in the single-source experiments of \cite{DiorioJFM} or in the present results when only one air jet was turned on. 

\begin{figure}
\begin{center}
\begin{tabular}{cccc}
  (\textit{a}) & (\textit{b}) & (\textit{c}) & (\textit{d})\\
  \includegraphics[height=0.95in]{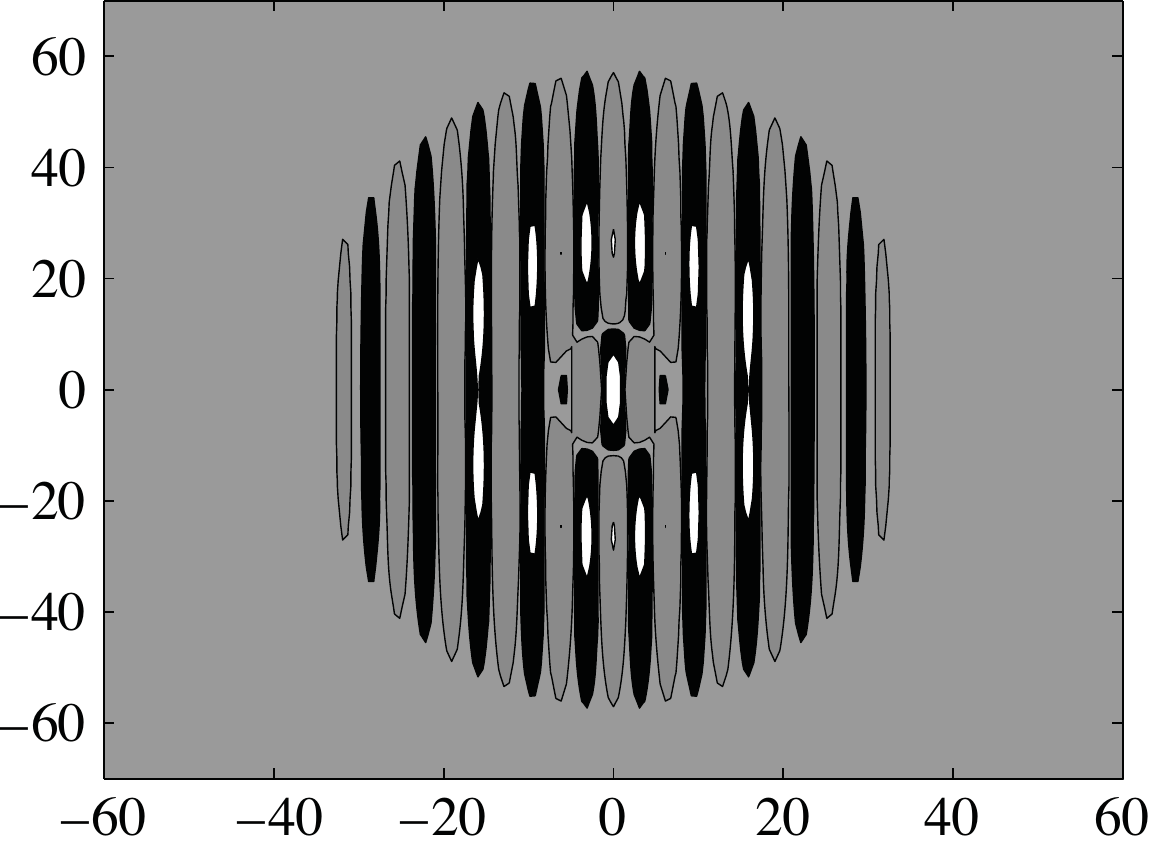}&
  \includegraphics[trim=0.25in 0 0 0,clip=true,height=0.95in]{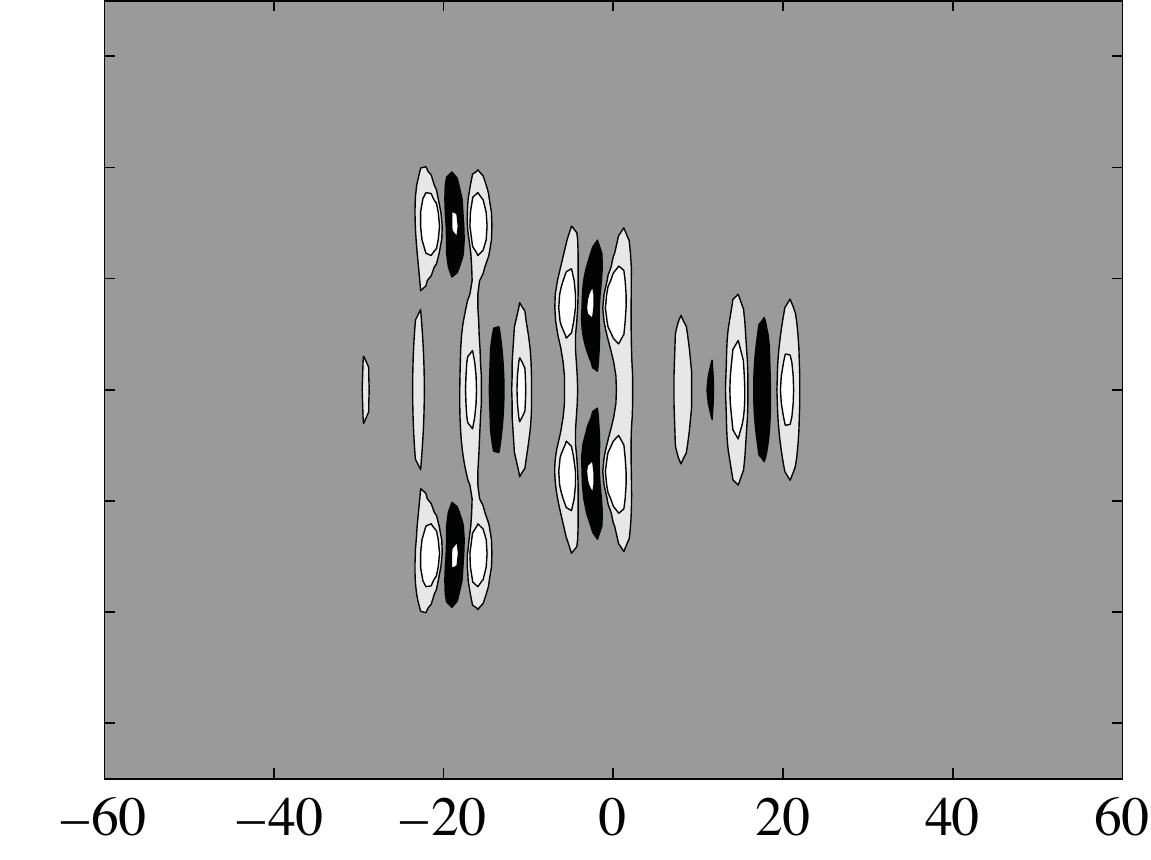}&
  \includegraphics[trim=0.25in 0 0 0,clip=true,height=0.95in]{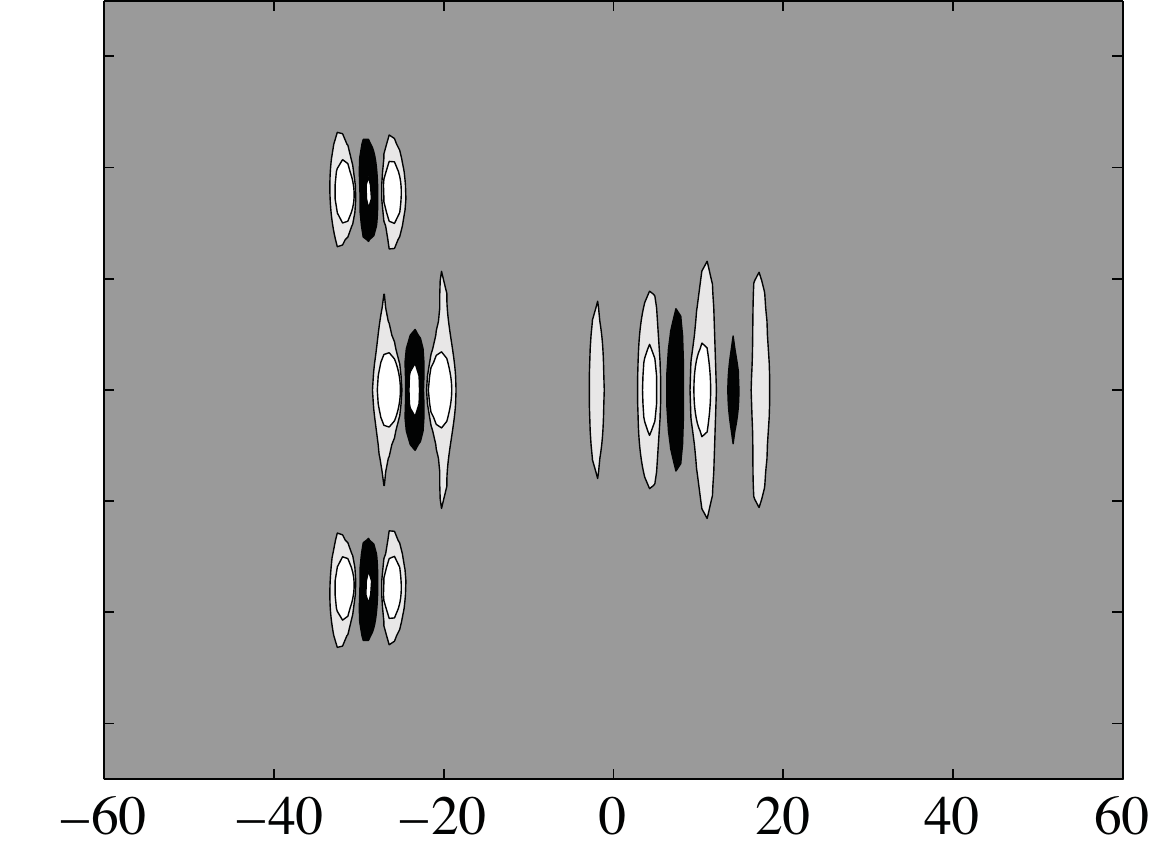}&
  \includegraphics[trim=0.25in 0 0 0,clip=true,height=0.95in]{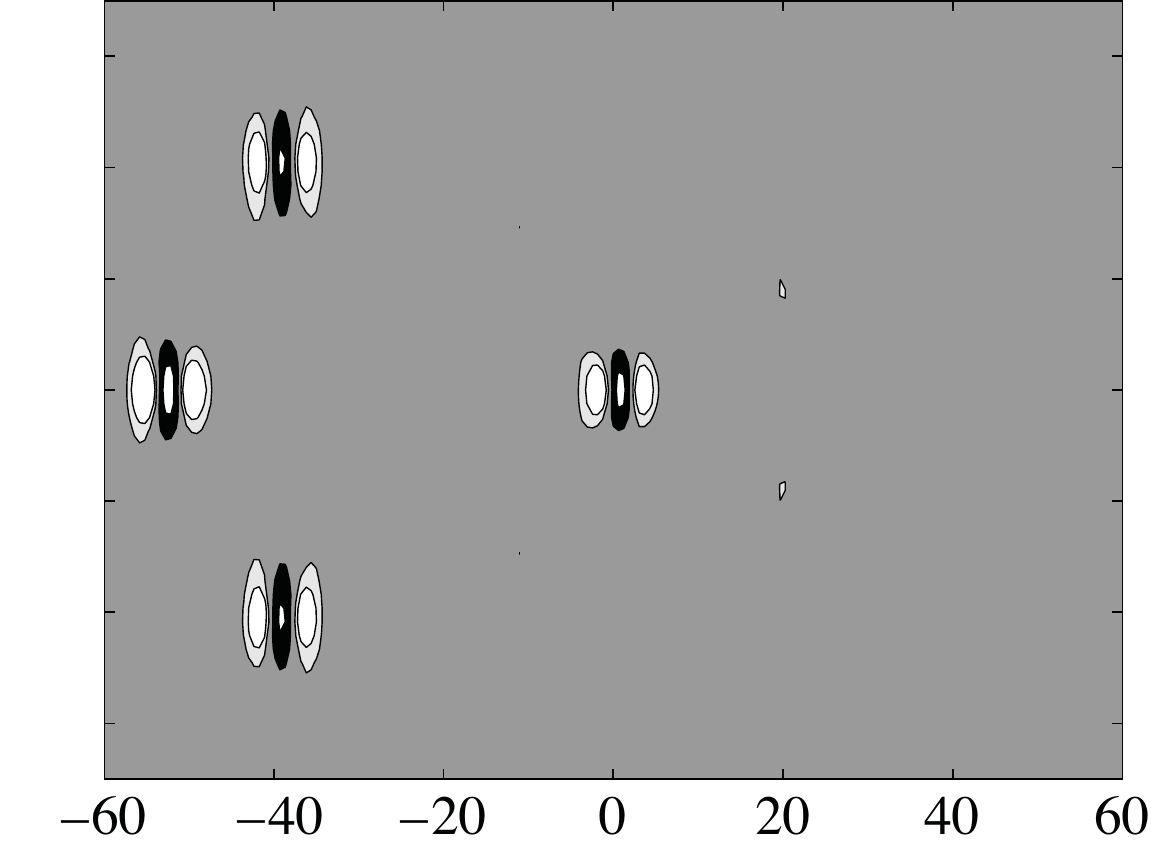}\\
\end{tabular}
\end{center}
\caption{Calculations of the evolution of a complex solitary wave  into several depression breathers. From figure 12 in \cite{Milewski2012}. }
\label{fig:Wang}
\end{figure}

Similar patterns of quasi-steady lumps were also found in  the inviscid calculations of \cite{Milewski2012}, where the evolution of a complex solitary wave that appears as a group of 12 lumps forming a ring around a central lump was examined.  This nonlinear pattern evolved into a group of four depression breathers in three rows containing one, two and one breathers, see figure~\ref{fig:Wang}.  The successive  rows of one an two lumps in images (\textit{b}) to (\textit{d}) are remarkably similar to the mean patterns found in the present experiments.  

\begin{figure}
\begin{center}
\begin{tabular}{cc}
\vspace{0.01in} &  (\textit{c})\\
\vspace{0.01in} & \multirow{2} {*} {\includegraphics[trim=0 0 0 0.6in,clip=true,width=2.25in]{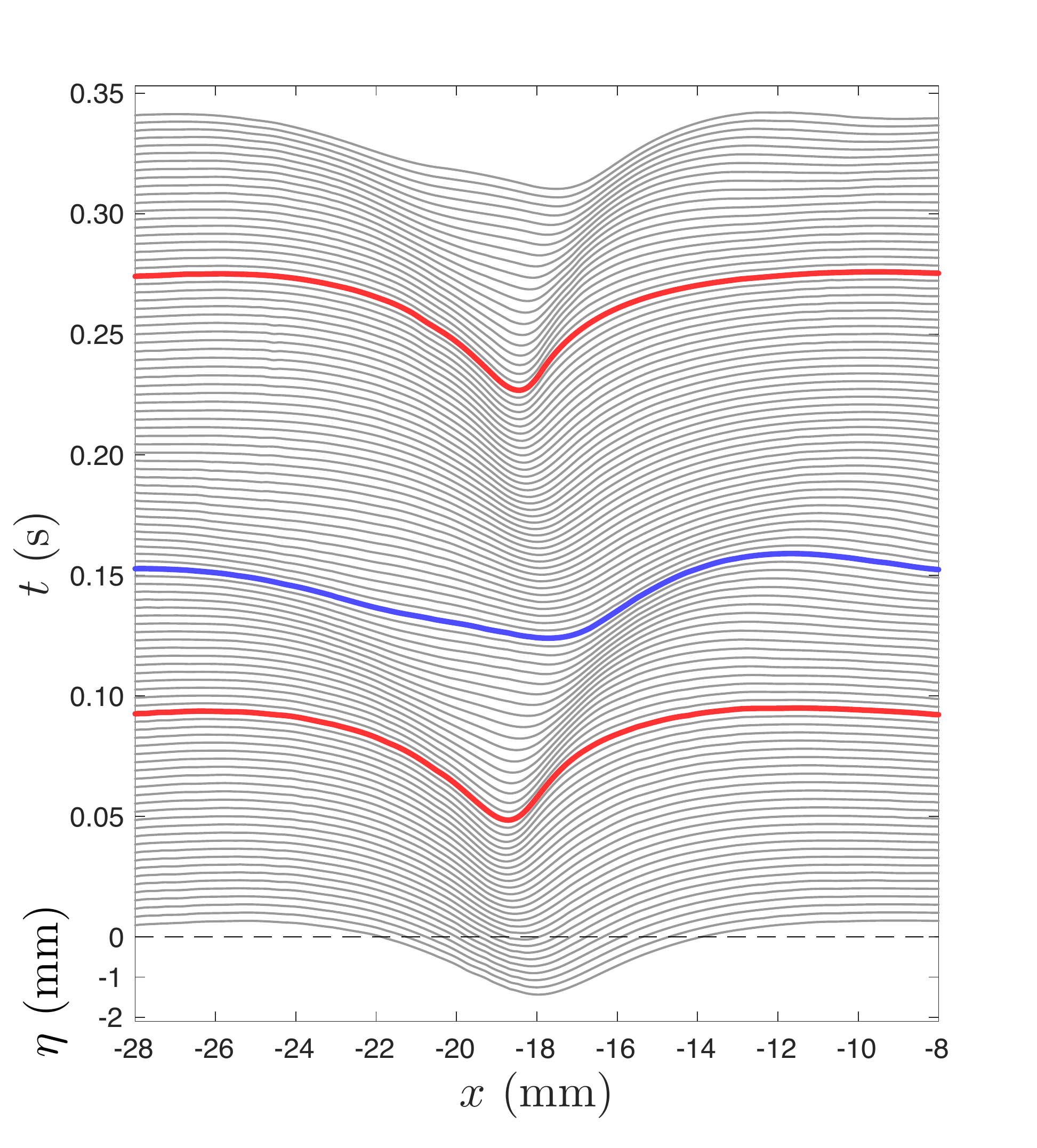}}\\
(\textit{a}) \vspace{0.05in}\\
\includegraphics[width=2.3in]{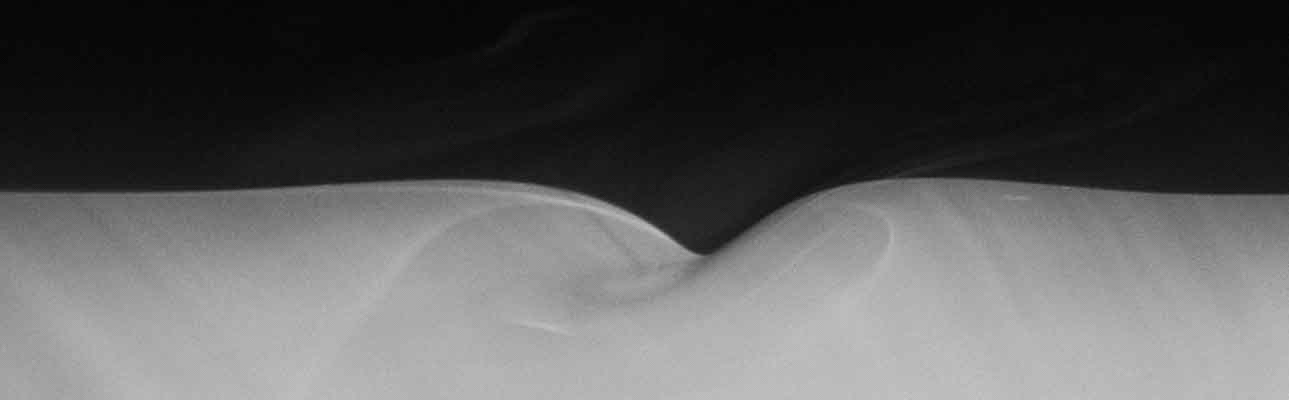}\\
(\textit{b}) \vspace{0.05in}\\
\includegraphics[width=2.3in]{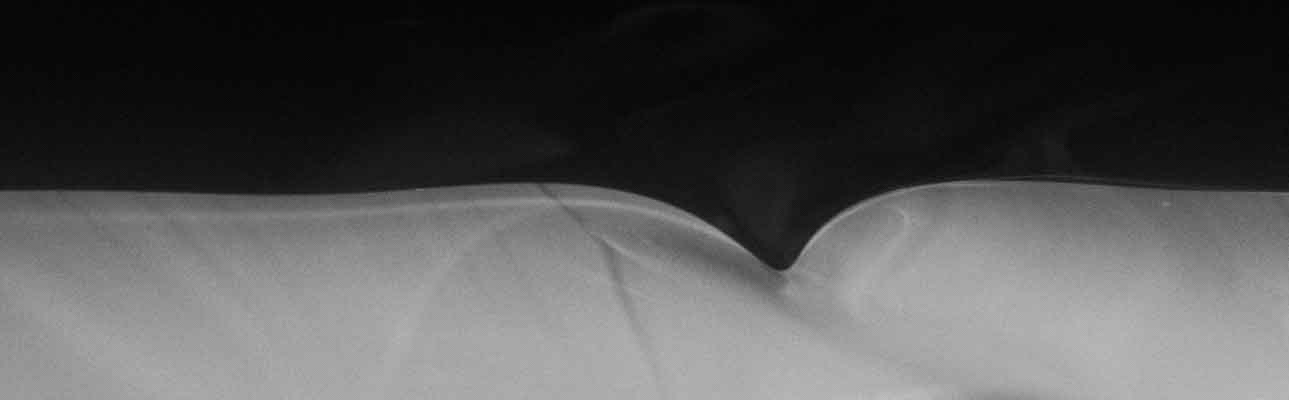}
\end{tabular}
\end{center}
\vspace{0.3in}
\caption{LIF profile measurements along the center-plane of the tubes with $\delta_{T}=4.7$. The sources are moving from left to right. (\textit{a}) LIF image when only one air jet is on. (\textit{b}) LIF image when both air jets are on. (\textit{c}) Evolution of the free surface profiles  when both air jets are on. Each profile is shifted vertically by 0.2~mm from the previous profile and the time difference between consecutive profiles is 1/300 s.}
\label{fig:LIF_raw}
\end{figure}

The LIF images are used to obtain quantitative measurements of the wave height along the midline between the two tubes. Examples of raw LIF images are shown in figure \ref{fig:LIF_raw} (also see movie 5). In these images the source is moving from left to right. Figure \ref{fig:LIF_raw}(\textit{a}) shows a state III lump (only one air jet is on) as it passes through the laser light sheet. It is emphasized that this lump is oriented with the normal to its long axis making an angle of approximately $15\degree$ to the streamwise direction. Figure \ref{fig:LIF_raw}(\textit{b}) shows the surface shape when both air jets are on at the instant just before the first burst. In this case, the normal to the long axis of the depression is in the streamwise direction, i.e. in the plane of the light sheet. Both images are taken at the same instant relative to the start of the carriage motion, in separate runs with the identical carriage motions. In the case when both air jets are on, the depression in the midline of the air-jet tubes becomes asymmetric and very steep (almost vertical at the front edge) just before the burst event.  

\begin{figure}
\begin{center}
\begin{tabular}{cc}
(\textit{a}) & (\textit{b})\\
  \includegraphics[trim=0.05in 0 0.7in 0in,clip=true,height=2.0in]{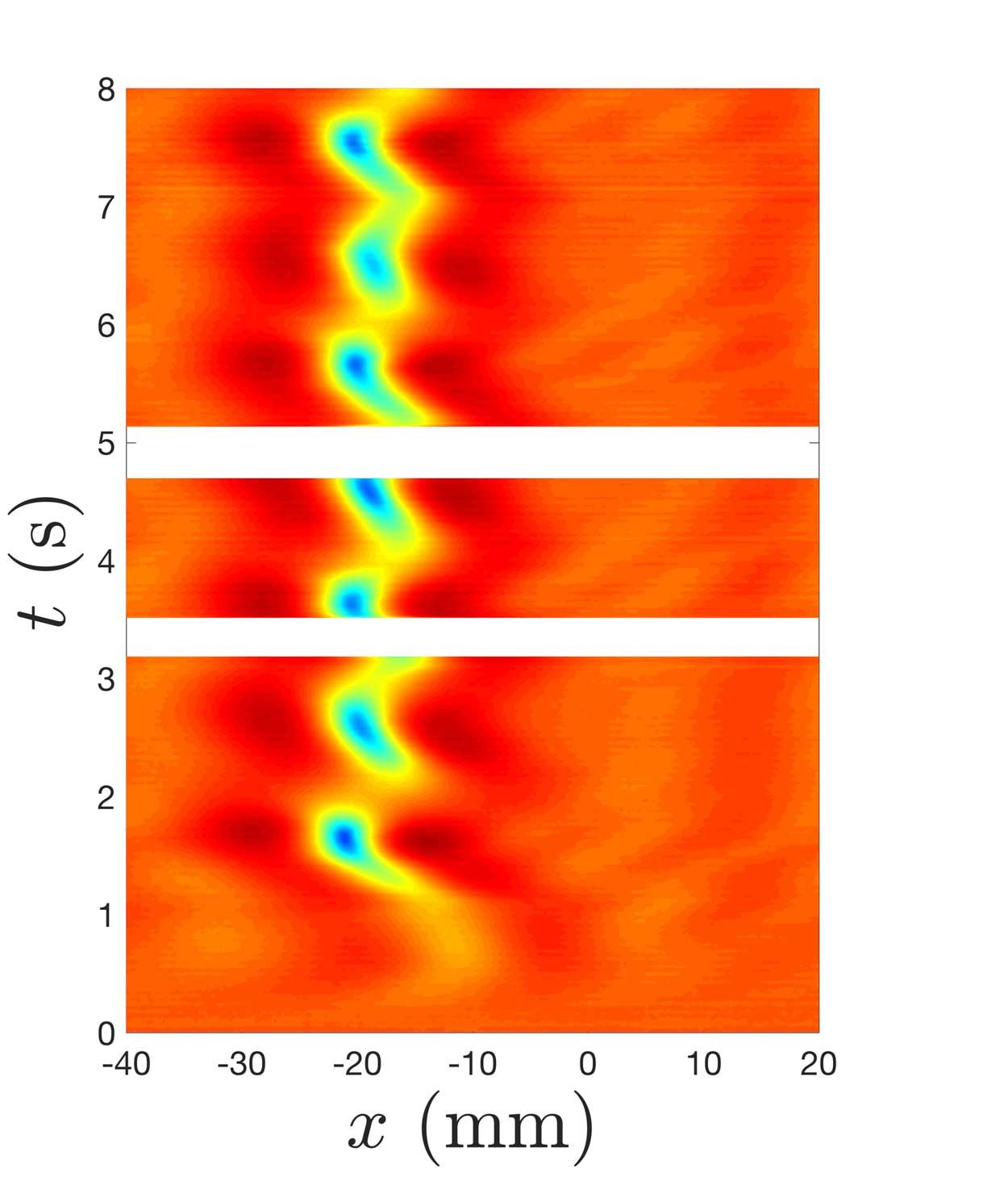}&
  \includegraphics[trim=0.4in 0 0.7in 0in,clip=true,height=2.00in]{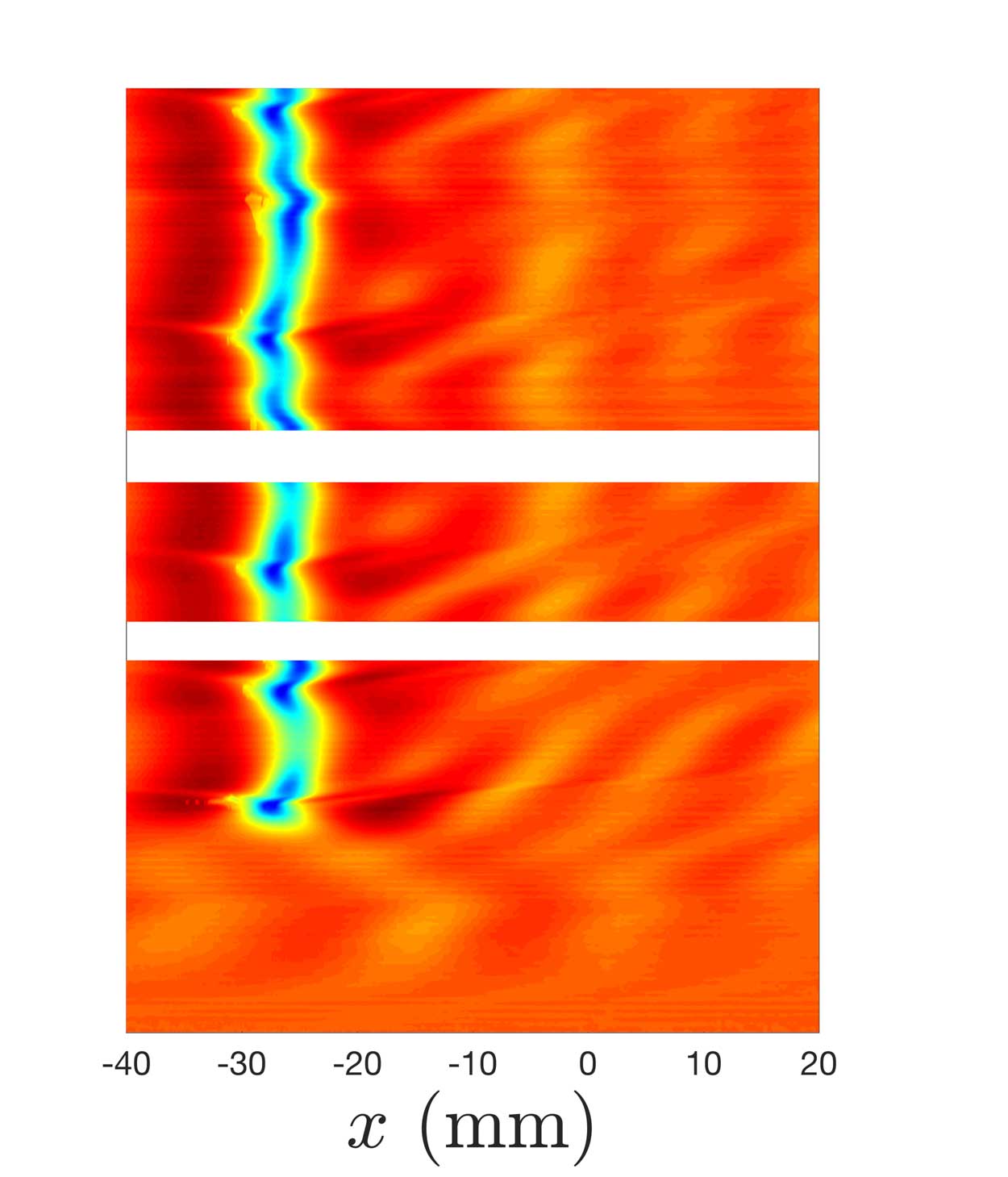}\\
  \includegraphics[trim=0in 0 0.8in 0,clip=true,height=0.8in]{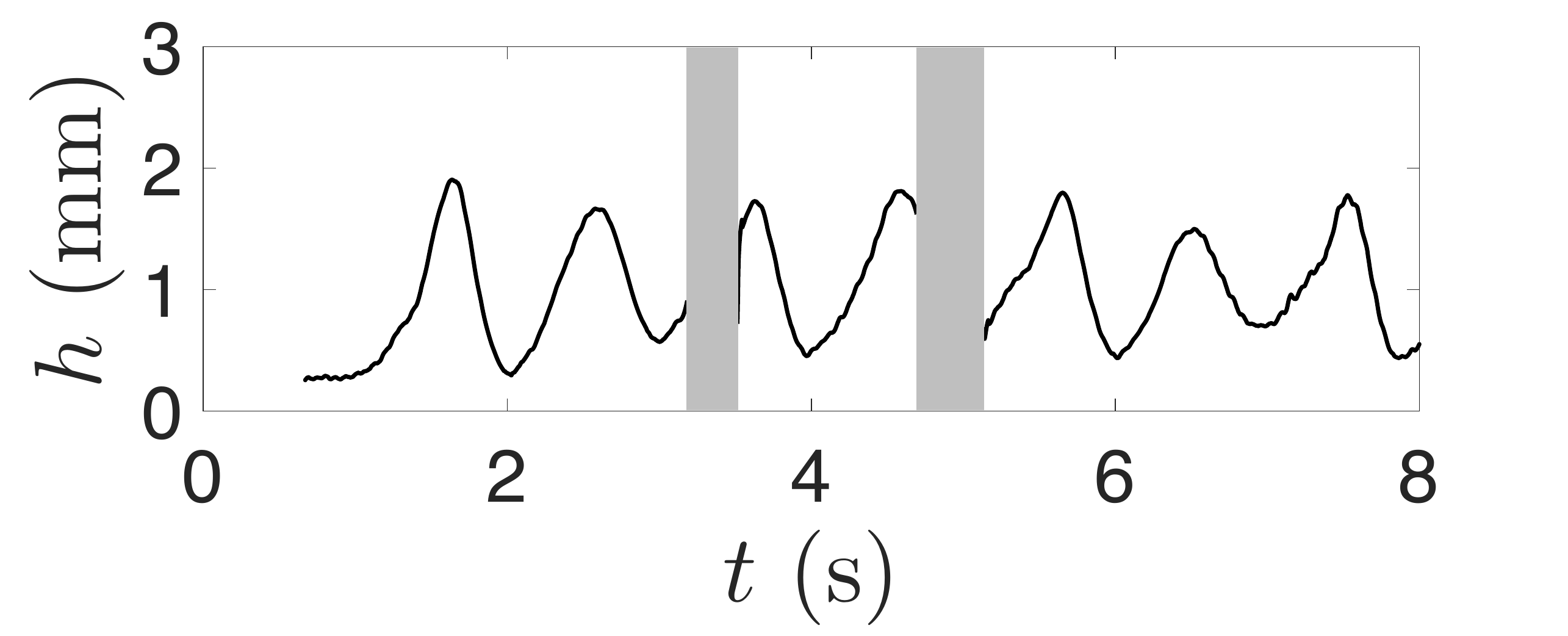}&
  \includegraphics[trim=1.2in 0 0.8in 0,clip=true,height=0.8in]{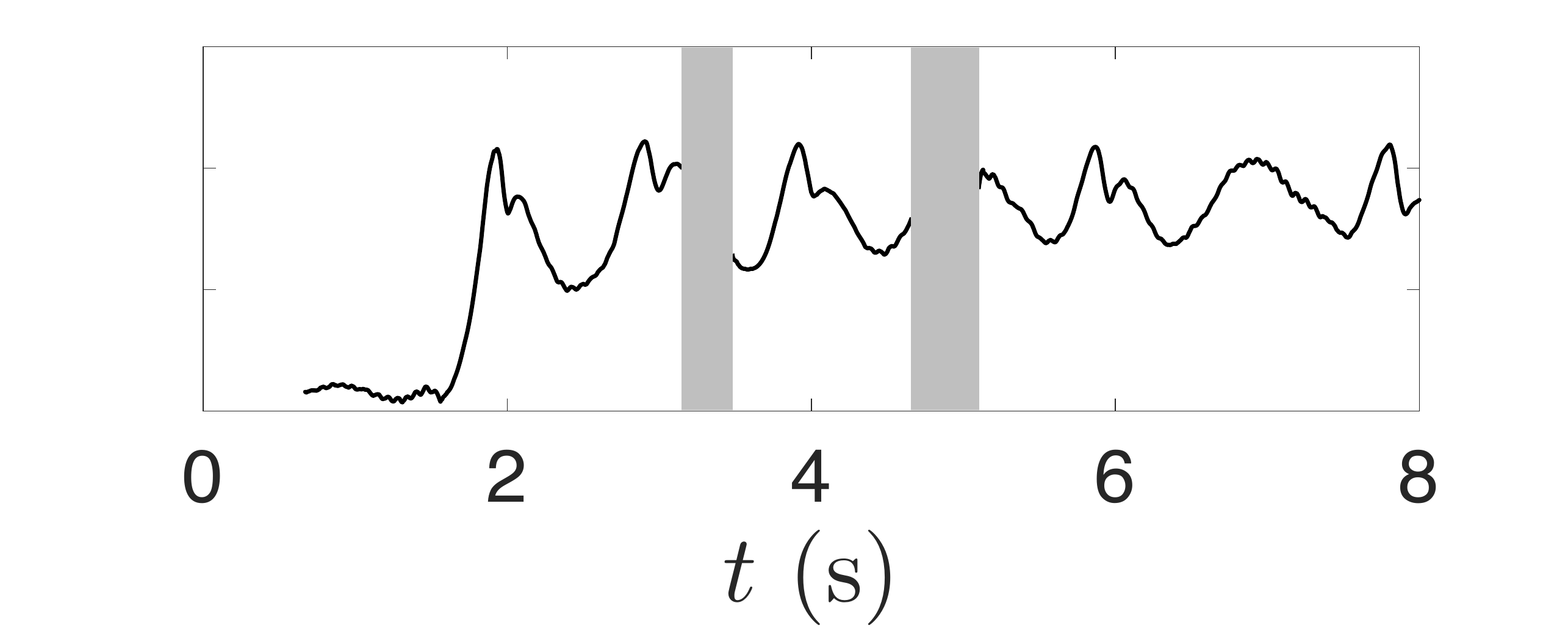}\\
(\textit{c}) & (\textit{d})\\
  \includegraphics[trim=0.05in 0 0.7in 0in,clip=true,height=2.0in]{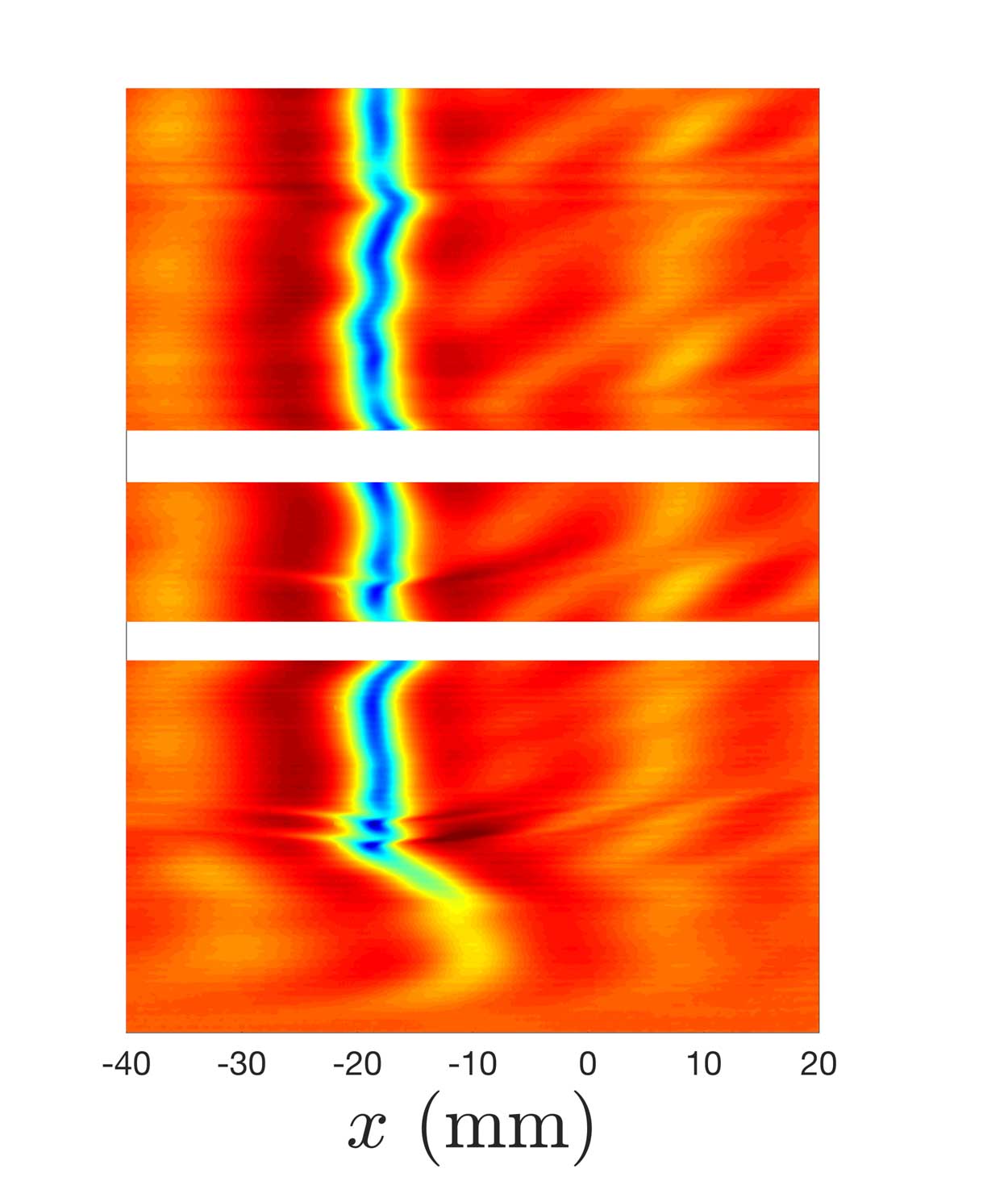}&
  \includegraphics[trim=0.4in 0 0.19in 0in,clip=true,height=2.0in]{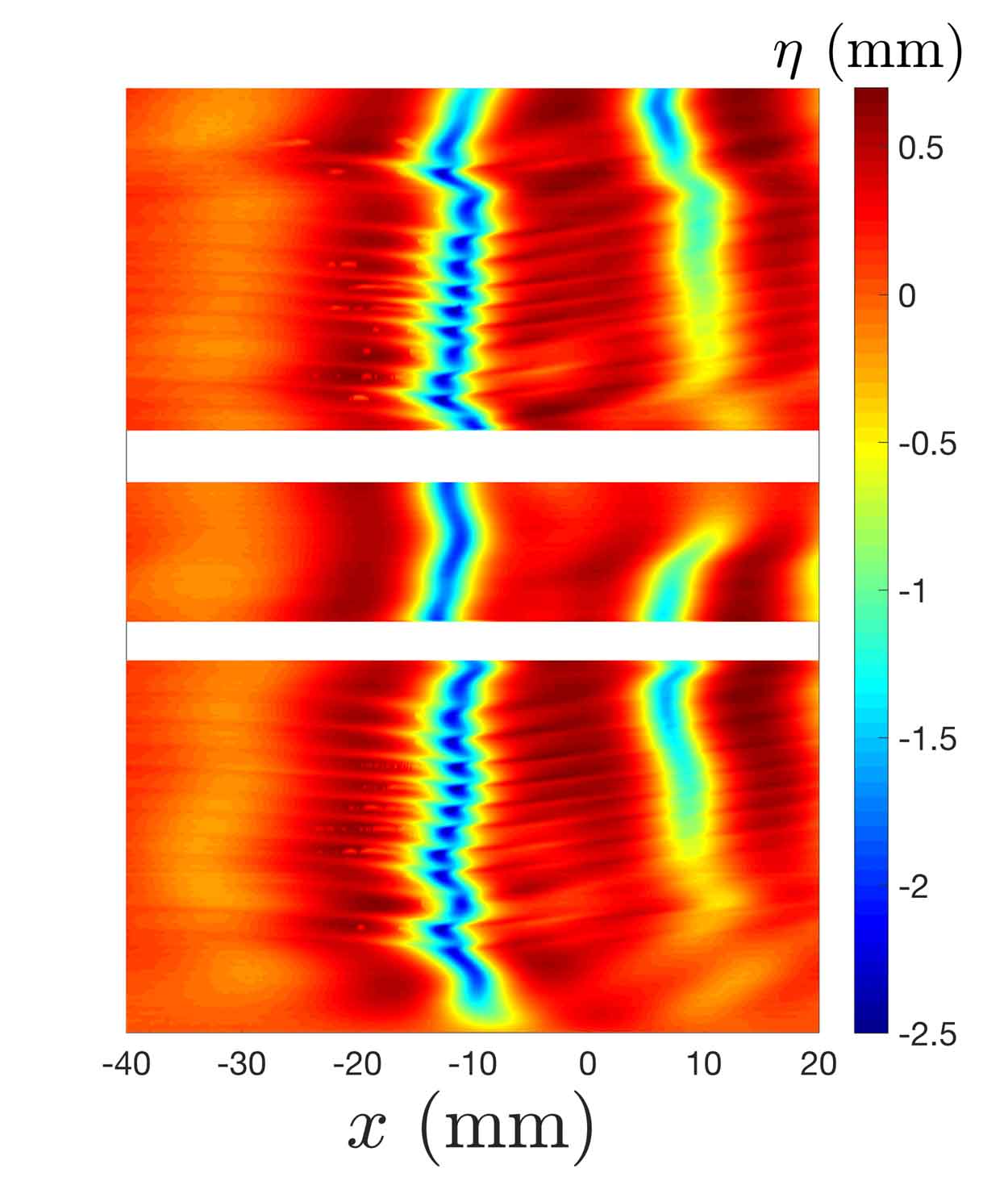}\\
  \includegraphics[trim=1.2in 0 0.8in 0,clip=true,height=0.8in]{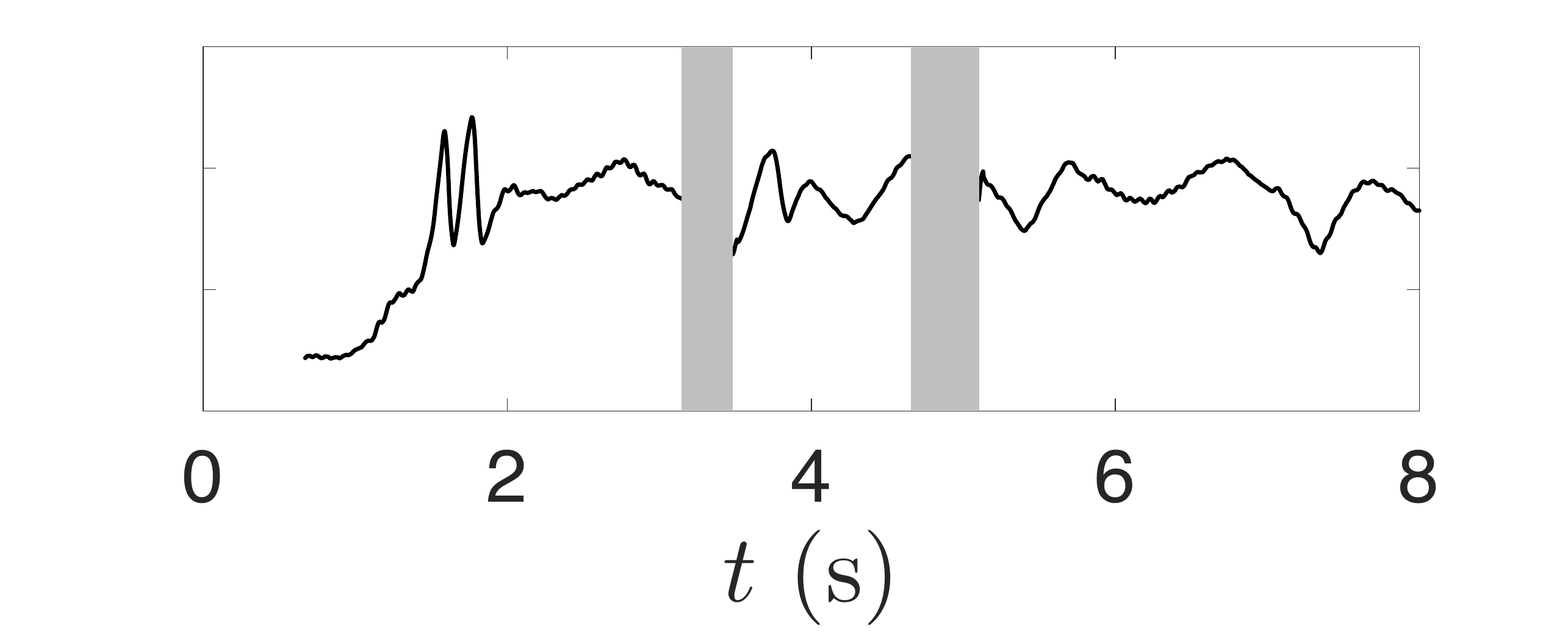}&
  \includegraphics[trim=1.2in 0 0in 0,clip=true,height=0.8in]{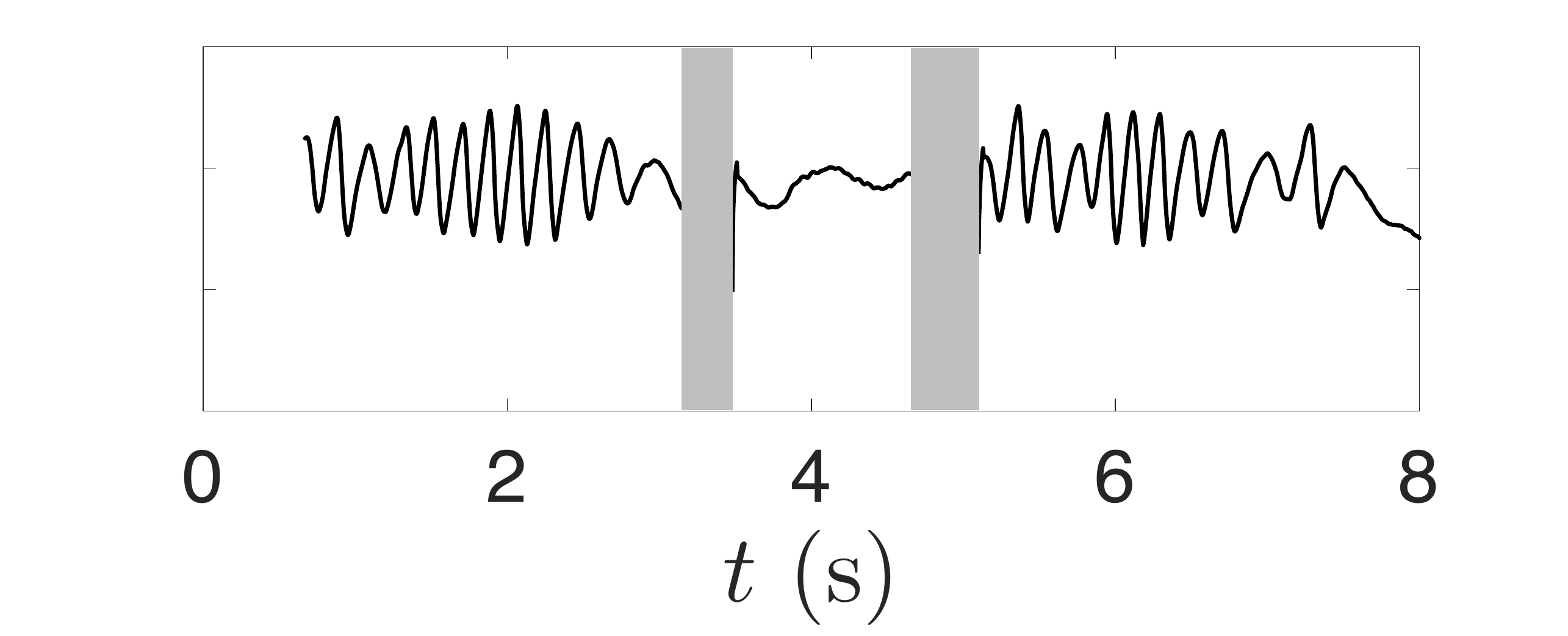}\\
\end{tabular}
\end{center}
\caption{Top row: surface height in the $x$-$t$ plane. Bottom row: maximum depth vs. time. (\textit{a}) Only one air jet active. (\textit{b}) $\delta_{T}=7.0$. (\textit{c}) $\delta_{T}=4.7$. (\textit{d}) $\delta_{T}=2.3$.}
\label{fig:D_effect}
\end{figure}

In order to illustrate the evolution of the free surface shape during the collision of the lumps, a sequence of profiles extracted from the LIF images for $\delta_{T}=4.7$ is shown in figure \ref{fig:LIF_raw}(\textit{c}). In this figure, the sources are moving in the positive x-direction and are located at $x=0$. 
Each profile is shifted vertically from the previous profile to show the evolution in time. 
The red (blue) lines denote the profiles for which the maximum depth of the wave is a local maximum (minimum) in time. 
This set of profiles shows the first collision of the lumps and the subsequent radiation of small-amplitude waves, which are shown more clearly in the contour plots in figure~\ref{fig:D_effect} as discussed below. The surface profile of the local depression resembles the stream-wise profile of a gravity-capillary lump but becomes asymmetric and very steep (especially the front edge) as the two lumps approach the mid-plane. As the energy from the sources accumulates and the depth of the depression increases, it moves downstream, consistent with the fact that the speed of a gravity-capillary lump decreases with increasing depth. Eventually this depression loses its form and radiates energy away as small-amplitude radial waves (between the red and blue curves in the plot). Shortly after this burst, the energy accumulates again and another burst occurs. The period of this fast oscillation is 0.18~s, which is five times smaller than the shedding period of lumps for one pressure source with the same $\epsilon$ and $\alpha$. This phenomenon of fast oscillations accompanied by radiation of small-amplitude waves is typical of the first collision of lumps for all tube separations used in the experiments.
The average speed of this depression is 0.994$c_{min}$, i.e. the towing speed $U$.  According to the fully nonlinear potential flow steady state calculations of 
E. P\u{a}r\u{a}u (personal communications 2016), the maximum depth of the depression would be 0.52~mm while the depth here reaches about 2.2~mm just prior to the burst.  In between the double burst events, the depression diminishes to only 1.1~mm.  A depth of 2.2~mm corresponds to a speed lower than those calculated by E. P\u{a}r\u{a}u.

Figure \ref{fig:D_effect} shows plots of the time evolution of the center-line surface profiles for a case with one air jet (column \textit{a}), and cases with two air jets separated by $\delta_{T}=7.0$ (\textit{b}), 4.7 (\textit{c}), and 2.3 (\textit{d}). The top row consists of contour plots of  the surface height profile encoded as color on an $x$-$t$ plane. The air-jet tubes are approximately located at $x=0$. The bottom row of plots show the maximum depth of each surface height profile versus time. In the plots in column  (\textit{a}), the LIF light sheet was located 4~cm laterally from the air-jet tube.

When one air jet is on (column \textit{a}), large single depressions appear at $x\approx -22$~mm with a period of about 1~s.  These are the lumps that are shed in state III and then travel through the light sheet.  With both air jets on and $\delta_{T}=7.0$ (column \textit{b}), maximum depressions again appear with a 1-second period.  This supports the idea that when the distance between the two sources is large, they radiate energy with a periodicity equal to that of lump generation with a single source, in spite of the fact that the pattern with two jets appears as a quasi-steady array of lump-like depressions after the initial lump impact.  Each maximum depression is followed by a slightly smaller local maximum depression about 0.1~s later.  This double maximum is associated with the radiation of smaller-amplitude waves.  Since the profile measurements are taken in the reference frame of the air-jet tubes, features in the contour plots that appear as vertical lines are maintaining their positions relative to the tubes. Thus, for the $\delta_{T}=7.0$ case, it can be seen that the depression moves upstream slightly during each radiation event, only to fall back again as the next depression forms. Upstream of the depressions, the small-amplitude radiated waves can be seen as yellow regions with positive slope. When $\delta_{T}=4.7$, (column \textit{c}), a pattern similar to that for $\delta_{T}=7.0$ is found, but in this case the maximum depression is nearly continuous in time, except for several incidents of the double maximum and wave radiation at $t\approx1.8$ and 3.8~s. The strongest of these double oscillations is the first one and its period is again about 0.1~s. The time of this first oscillation is close to that in the single source case, where the LIF light sheet was also 4~cm away from the source. When $\delta_{T}=2.3$, the initial depression occurs even earlier and initiates a rapid oscillation of the maximum depression height that lasts for about 12 cycles.  This is followed by a calm period and then another period of oscillations. The oscillation period is about 0.1~s in all cases.

\section{Conclusions}

The interaction of gravity-capillary lumps generated by two surface pressure sources moving side by side at a constant speed a little less than the minimum phase speed of linear gravity-capillary waves was investigated experimentally. When a single source moves at this speed, an unsteady V-shaped wave pattern appears behind the source and sheds gravity-capillary lumps from the tips of the V  \cite[]{DiorioPRL} at a period of about 1~s.  With two sources, as they start from rest,  lumps that are generated simultaneously from the near sides of each V meet and interact downstream at the center-plane between the two sources. The collision of the first pair of lumps results in  a steep depression that radiates small-amplitude waves.  When the separation between the two sources is relatively large ($7.0\lambda_{min}$, where $\lambda_{min}$ is the wavelength of the linear wave with minimum phase speed), this initial collision results in a two-cycle 0.1-s-period oscillation of the depression at the collision site.   The original depressions do not pass through the collision site, i.e. there is no evidence of soliton-like collision behavior.  Subsequent to the first collision, a quasi-steady pattern of three rows of depression pairs forms in the streamwise region between the central depression and the sources.  Once formed, this depression pattern undergoes a breather-type oscillation with a period about equal to the shedding period for a single source (1~s).  During this oscillation, the downstream depression on the midline of the pattern periodically becomes very steep and generates a burst of small-amplitude waves.  
This behavior indicates that, in spite of the relatively fixed location of the depressions, energy continues to radiate periodically from the sources to the central depression.  As the separation between the sources is decreased, the generation of small-amplitude waves during the initial collision persists, but  the number of rows of depressions decreases and the breather oscillation becomes erratic.  For the smallest source separation studied herein ($2.3\lambda_{min} $), the fast oscillation cycle of the center depression dominates and continues for many cycles, sometimes stopping and then resuming a short time later.  More thorough experiments and theoretical models are needed to provide a physical explanation for the observed phenomenon.\bigskip

This work was partially supported by the Office of Naval Research under grant N000141110029  and the National Science Foundation, Division of Ocean Sciences under grant OCE0751853. 

\bibliography{Collision}
\bibliographystyle{jfm}

\end{document}